\documentclass[11pt]{article}

\usepackage{amsmath}
\usepackage{amssymb}
\usepackage{latexsym}
\usepackage{amsmath, amsfonts,amssymb, amsthm, euscript,makeidx,color,mathrsfs}
\usepackage{xcolor}
\usepackage{indentfirst}
\usepackage{algpseudocode}
\usepackage{graphicx}
\usepackage{epstopdf}
\usepackage{float}
\usepackage{subfig}
\usepackage{hyperref}

\newtheorem{assumption}{Assumption}
\def\qed{ \ \vrule width.2cm height.2cm depth0cm\smallskip}

\newcommand{\ba}{\begin{array}}
\newcommand{\ea}{\end{array}}
\newcommand{\be}{\begin{equation}}
\newcommand{\ee}{\end{equation}}
\newcommand{\bea}{\begin{eqnarray}}
\newcommand{\eea}{\end{eqnarray}}
\newcommand{\beaa}{\begin{eqnarray*}}
\newcommand{\eeaa}{\end{eqnarray*}}

\def\neg{\negthinspace}

\def\a{\alpha}

\def\e{\varepsilon}

\def\m{\mu}

\def\si{\sigma}

\def\o{\omega}

\def\O{\Omega}

\def\Th{\Theta}

\def\Th{\Theta}

\def\O{\Omega}
\def\cF{{\cal F}}
\def\cG{{\cal G}}

\def\hB{\mathbb{B}}
\def\hC{\mathbb{C}}
\def\hD{\mathbb{D}}
\def\hE{\mathbb{E}}
\def\hF{\mathbb{F}}

\def\hL{\mathbb{L}}
\def\hM{\mathbb{M}}
\def\hN{\mathbb{N}}

\def\hP{\mathbb{P}}

\def\hR{\mathbb{R}}
\def\hS{\mathbb{S}}
\def\hT{\mathbb{T}}
\def\hU{\mathbb{U}}
\def\hV{\mathbb{V}}

\def\hX{\mathbb{X}}

\def\sA{\mathscr{A}}

\def\sC{\mathscr{C}}

\def\sE{\mathscr{E}}

\def\sG{\mathscr{G}}
\def\sH{\mathscr{H}}

\def\sK{\mathscr{K}}
\def\scL{\mathscr{L}}

\def\sP{\mathscr{P}}

\def\no{\noindent}

\def\ss{\smallskip}
\def\ms{\medskip}

\def\q{\quad}
\def\qq{\qquad}

\def\bm{{\bf m}}

\def\tr{\hbox{\rm tr}}

\def\qed{ \hfill \vrule width.25cm height.25cm depth0cm\smallskip}

\newcommand{\basa}{\begin{assumption}}
\newcommand{\easa}{\end{assumption}}

\newcommand{\ol}{\overline}

\newcommand{\bas}{\begin{assum}}
\newcommand{\eas}{\end{assum}}

\def\essinf{\mathop{\rm essinf}}

\def\co{\mathop{{\rm co}}}

\def\tr{\hbox{\rm tr$\,$}}

\def\ol{\overline}

\def\cl{\mbox{\rm cl}}
\def\co{\mbox{\rm co}}

\def\dis{\displaystyle}

\def\bi{{\bf i}}

\def\1{{\bf 1}}

\def\:{\!:\!}
\def\reff#1{{\rm(\ref{#1})}}

\DeclareMathOperator{\dec}{dec}

\DeclareMathOperator{\gr}{graph}
\DeclareMathOperator{\diag}{diag}

at 9pt

\newcommand{\of}[1]{\ensuremath{\left( #1 \right)}}
\newcommand{\cb}[1]{\ensuremath{ \left\{ #1 \right\} }}
\newcommand{\sqb}[1]{\ensuremath{ \left[ #1 \right] }}
\newcommand{\norm}[1]{\ensuremath{ \left\Vert #1 \right\Vert }}
\newcommand{\ip}[1]{\ensuremath{ \left\langle #1 \right\rangle }}

\def\prehp(#1,#2){\ensuremath{  #1 \cdot #2 }}

\begin{document}

\newtheorem{thm}{Theorem}[section]
\newtheorem{lem}[thm]{Lemma}
\newtheorem{cor}[thm]{Corollary}
\newtheorem{prop}[thm]{Proposition}
\newtheorem{rem}[thm]{Remark}
\newtheorem{eg}[thm]{Example}
\newtheorem{defn}[thm]{Definition}
\newtheorem{assum}[thm]{Assumption}

\renewcommand {\theequation}{\arabic{section}.\arabic{equation}}
\def\thesection{\arabic{section}}

\title{\bf Superhedging under Proportional Transaction Costs in Continuous Time}
\author{
Atiqah Almuzaini\thanks{\noindent {Department of Mathematics and Statistics, University of Jeddah, Jeddah, Saudi Arabia.
Email: ahalmuzaini@uj.edu.sa}}, ~
 \c{C}a\u{g}{\i}n Ararat\thanks{Department of Industrial Engineering, Bilkent University, Ankara,~06800, Turkey. Email: cararat@bilkent.edu.tr. This author is supported by T{\"{U}}B\.{I}TAK~2219 Program and by the Fulbright Scholar Program of the U.S. Department of State, sponsored by the Turkish Fulbright Commission. This work was partly completed while the author was visiting University of Southern California, whose hospitality is greatly appreciated.},
~ and ~ Jin Ma\thanks{ \noindent Department of
Mathematics, University of Southern California, Los Angeles, 90089;
email: jinma@usc.edu. This author is supported in part by US NSF grants \#2205972 and \#2510403.
}} 
\date{November 22, 2025}
\maketitle

\begin{abstract}
We revisit the well-studied superhedging problem under proportional transaction costs in continuous time using the recently developed tools of set-valued stochastic analysis. By relying on a simple Black-Scholes-type market model for mid-prices and using continuous trading schemes, we define a dynamic family of superhedging sets in continuous time and express them in terms of set-valued integrals. We show that these sets, defined as subsets of Lebesgue spaces at different times, form a dynamic set-valued risk measure with multi-portfolio time-consistency. Finally, we transfer the problem formulation to a path-space setting and introduce approximate versions of superhedging sets that will involve relaxing the superhedging inequality, the superhedging probability, and the solvency requirement for the superhedging strategy with a predetermined error level. In this more technical framework, we are able to relate the approximate superhedging sets at different times by means of a set-valued Bellman's principle, which we believe will pave the way for a set-valued differential structure that characterizes the superhedging sets.
\end{abstract}

\vfill


\newpage

\date{}
\maketitle

\section{Introduction }
\setcounter{equation}{0}

In this paper, we revisit the well-known dynamic superhedging problem in continuous time under proportional transaction costs and offer a new approach for it through the lens of some recently studied tools of set-valued stochastic analysis.

\ms
\no{\bf Motivation and Literature Review.} In incomplete financial markets, given an uncertain claim, one may look for initial endowments that will yield a better terminal payoff than the given claim under all scenarios. Each such endowment is said to superhedge the claim and the superhedging problem aims to find the ``cheapest" such endowment. Our focus in this paper is on markets with transaction costs as a particular case of incompleteness, which has been studied in the literature at least for three decades. Among the earliest works, Jouini and Kallal \cite{jouini} studied the absence of arbitrage (or free lunch) and provided a dual characterization of the superhedging price in discrete-time and continuous-time frameworks with multiple underlying assets. Within the same stream of research, Cvitani\' c and Karatzas \cite{cvi} studied a more concrete model with a single risky asset and a riskless asset in continuous time; one of their main results is a formula for the superhedging price using dual supermartingale measures.

In 1999, Kabanov \cite{Kcurrency} introduced a new framework for multi-asset markets with proportional transaction costs. By considering currency markets as the canonical example of these markets, contingent claims are modeled in terms of \emph{physical units}, hence as random vectors in this framework. The transaction costs are encoded by a so-called \emph{solvency cone}, which consists of portfolio vectors that can be exchanged into portfolios with long positions in all assets. Accordingly, in the superhedging problem, one now looks for the \emph{superhedging set}, i.e., the set of \emph{all} portfolio vectors as initial endowments that can superhedge the claim. In this framework, dual representation theorems for the superhedging set were proved by Kabanov et al.~\cite{KabanovRasonyi}, Schachermayer \cite{Sch} in discrete time; by Kabanov and Last \cite{KandG}, Kabanov and Stricker \cite{KandS}, Campi and Schachermayer \cite{Campi} in continuous time under different levels of generality; see also the book by Kabanov and Safarian \cite{KabanovSafarian} for an extensive treatment of the subject.

The superhedging set considered in the references mentioned above is static, i.e., it consists of deterministic portfolio vectors \emph{at time zero}. Naturally, one can also consider a conditional superhedging set at each point in time and obtain a dynamic family of sets of random vectors. This is indeed one of the prominent examples of dynamic \emph{set-valued risk measures} in discrete time as studied by Feinstein and Rudloff \cite{FR13}. A key feature of these sets is that they satisfy a set-valued backward recursion, also called \emph{multi-portfolio time-consistency}. Thanks to this property, as shown in \cite{lohnerudloff}, it is possible calculate these sets by a sequence of vector optimization problems when the underlying probability space is finite. More recently, the dynamic superhedging problem in discrete time is revisited in \cite{araratfeinstein}, where dynamic set-valued risk measures with multi-portfolio time-consistency, with superhedging sets being an example, are characterized as both the reachable sets of stochastic difference inclusions and the solutions of set-valued backward stochastic difference equations.

In continuous time, beyond trivial cases, it is an open problem to characterize set-valued dynamic risk measures in terms of backward stochastic differential structures such as inclusions and set-valued equations. One of the major obstacles in having such a characterization is that the literature on set-valued stochastic analysis largely focuses on \emph{forward} differential structures with \emph{bounded} sets (cf. \cite{K,MK}). Forwardness makes the stochastic analysis easier, e.g., it avoids issues related to the stochastic integral representation of set-valued martingales; see \cite{amw} for a recent well-posedness result for backward stochastic differential equations with bounded sets. Boundedness makes the set-valued analysis easier due to the convenient framework provided by the Hausdorff metric; see \cite{AlMa} for a recent well-posedness of forward set-valued stochastic differential equations with unbounded sets having a special structure. More seriously, it is not clear what the limiting nature of the difference inclusions and set-valued difference equations in \cite{araratfeinstein} should be when one passes to continuous time.

In addition to the technical issues discussed above for a general treatment, there are not many examples of dynamic set-valued risk measures in continuous time. To the best of our knowledge, the dynamic set-valued entropic risk measure studied in \cite[Section~6.2]{FR2015} is the only set-valued risk measure in continuous time that is multi-portfolio time-consistent beyond the conditional expectation-based risk measure. However, both risk measures are known to reduce down to vector-valued dynamic functionals, hence they do not possess any challenging features of the set-valued case.

\ms
\no{\bf Main Contributions.} In this paper, we construct a family of superhedging sets as an example of a dynamic set-valued risk measure in continuous time. To keep the technical complications at a minimum, we consider a multi-asset generalized Black-Scholes model under proportional transaction costs. In particular, this model generalizes the one in \cite{cvi} with a single risky asset. In this model, we provide a detailed description of the \emph{solvency cone}. As pointed out in various works on superhedging (cf., e.g., \cite{Campi, KandG,KabanovSafarian}), the definition of an admissible trading strategy is not as straightforward as in discrete time but one can work with processes of bounded variation whose Radon-Nikodym derivative with respect to the total variation takes values in the solvency cone, hence jointly allowing for continuous and impulse trading. By taking advantage of our Black-Scholes-type model, we consider the simpler class of instantaneous trading schemes that are absolutely continuous with respect to the Lebesgue measure with a Radon-Nikodym derivative taking values in the solvency cone. This enables us to make more detailed calculations using It\^o formula at the vector-valued level and express the corresponding superhedging sets in terms of set-valued integrals. With these, we are able to prove that the superhedging sets form a dynamic set-valued risk measure that is multi-portfolio time-consistent. Hence, we obtain an example that is truly set-valued and is defined on a concrete market model.

More importantly, we show that the superhedging sets, as subsets of $\hL^2$ spaces indexed by time,  are related by a recursive relation that can be seen as a set-valued Bellman's principle, which has 
been key to analyze some multivariate dynamic problems in the recent literature. We refer the reader to \cite{kovacova} for the mean-risk problem in discrete time, to \cite{karnam2017dynamic} for controlled multi-dimensional backward stochastic differential equations, to \cite{dynamicgames} for dynamic games in discrete and continuous time, and to \cite{meanfield} for mean field games in continuous time.

Finally, by exploiting the decomposability of the superhedging sets in $\hL^2$, we associate a set-valued stochastic process to the family of superhedging sets of a fixed multidimensional claim $X$. However, as observed in earlier works dealing with multivariate dynamic programming in continuous time (cf.~\cite{dynamicgames,meanfield}) in game-theoretic settings, establishing a dynamic programming principle for such set-valued process seems to be a tall task. Instead, we will transfer the problem formulation to a path-space setting and introduce approximate versions of superhedging that will involve relaxing 1) the superhedging inequality, 2) the superhedging probability, and 3) the solvency requirement for the superhedging strategy with a predetermined error level $\varepsilon>0$. In this more technical framework, we will be able to relate the approximate superhedging sets at different times by means of a set-valued Bellman's principle. We conjecture that this principle can be used to obtain a set-valued differential structure that characterizes the superhedging sets. We leave this for future research as it requires a new set of analytical tools, such as set-valued differentiation or a set-valued Itô formula (e.g., as in \cite{hjb}), to be combined with Bellman's principle.

Our contributions can be summarized as follows:
\vspace{-2mm}
\begin{itemize}
	\item We provide a mathematically tractable formulation of the superhedging problem in continuous time using solvency cones and instantaneous trading strategies.
\vspace{-2mm}
\item We define superhedging sets as subsets of $\hL^2$ at different times using the notion of functional set-valued integral. We show that, modulo a sign change, these sets form a multi-portfolio time-consistent dynamic set-valued coherent risk measure. As a consequence of multi-portfolio time-consistency, we prove that a \emph{functional} set-valued dynamic programming principle holds for the superhedging sets.
\vspace{-2mm}
\item We prove a \emph{pathwise/local} version of the set-valued dynamic programming principle by introducing a formulation of the superhedging problem on the path space. This formulation is the continuous-time analog of the tree-based local formulation in discrete time studied in \cite{araratfeinstein}.
	\end{itemize}

The rest of this paper is organized as follows. In Section~\ref{sec:prelim}, we recall some basic definitions of set-valued analysis, set-valued risk measures, and the superhedging problem in discrete time. In Section~\ref{sec:market}, we introduce the multi-asset Black-Scholes-type market model together with its solvency cone, trading strategies, and portfolio processes. For completeness, we also study the dual of the solvency cone and its connection to consistent price processes. This is followed by two main sections on the superhedging problem in continuous time. In Section~\ref{sec:superhedging}, we investigate the functional formulation of the superhedging sets as subsets of the $\hL^2$ space and study their collection as a dynamic set-valued risk measure. In Section~\ref{sec:pathwise}, we formulate superhedging sets as set-valued random variables on the canonical path space and prove a dynamic programming principle for these sets that holds pathwise.

\section{Preliminaries}\label{sec:prelim}
\setcounter{equation}{0}

In this section, we set up the notation for the rest of the paper and recall some basic concepts about set-valued random variables and set-valued dynamic risk measures, with a final emphasis on the superhedging problem in discrete time.

Let $\hX$ be a separable real Banach space with norm $|\cdot|$. We equip $\hX$ with its Borel $\sigma$-field. For a set $A\subseteq \hX$, we write $\cl_{\hX} (A)$, $\co (A)$, $\overline{\co} (A)$ for the closure, convex hull, closed convex hull of $A$. We denote $\hB_{\hX}(\varepsilon)$ to be the closed norm ball in $\hX$ centered at the origin with radius $\varepsilon\geq 0$. When $\hX=\mathbb{R}^d$ is the Euclidean space with $d\in \hN:=\{1,2,\ldots\}$, we assume that $|\cdot|$ is the $\ell_2$-norm and we write $\cl(A)=\cl_{\hR^d}(A)$ for a set $A\subseteq \hX$. We denote $\sC(\hX), \sG(\hX)$ to be the families of all closed, closed convex nonempty subsets of $\hX$, respectively. For nonempty sets $A,B\subseteq \hX$ and $\lambda\in\hR$, we define the usual {\it Minkowski (elementwise) addition} and {\it multiplication by scalars} by
\[
A+B:=\{x+y\colon x\in A,\ y\in B\},\quad \lambda A:=\{\lambda x\colon x\in A\};
\]
we simply write $A+y=A+\{y\}$ for $y\in\hX$. We endow $\sC(\hX)$ with the {\it Hausdorff distance} $h$:  for $A,B\in \sC(\hX)$, we define
\[
h(A,B):=\max \{\overline{h}(A,B),\overline{h}(B,A) \}=\inf\{\e>0: A\subseteq B+\hB_{\hX}(\varepsilon),~B\subseteq A+\hB_{\hX}(\varepsilon)\}, 
\]
where $\overline{h}(A,B):=\sup\{ d(x, B)\colon x\in A\}$ and $d(x,B):=\inf\{|x-y|\colon y\in B\}$. It is well-known that $(\sC (\hX),h)$ is a complete metric space, allowing the value $+\infty$ for the metric $h$.

\ms
\no{\bf  Set-Valued Stochastic Analysis.} We recall some concepts in set-valued stochastic analysis, referring the reader to the well-known books \cite{K, MK} and the previous work \cite{AlMa} for more details.  To that end, let $(\O,\sE,\mu)$ be a finite measure space and denote $\hL^0_\sE(\O,\hR^d)$ to be the space of all $\sE$-measurable functions $f\colon\O\to\hR^d$ distinguished up to $\mu$-almost everywhere ($\mu$-a.e.) equality. A \emph{set-valued function} $F\colon\O\rightarrow \sC(\mathbb{R}^d)$ is said to be \emph{measurable} if, for each $A\in \sC(\mathbb{R}^d)$, it holds that $\{\o\in \O\colon F(\o)\cap A\neq \emptyset \}\in \sE$; we denote $\scL_{\sE}^0(\O,\sC (\mathbb{R}^d))$ to be the set of all such functions distinguished up to $\mu$-a.e. equality.
A function $f\in \hL^0_\sE(\O,\hR^d)$ is called a {\it selector} for $F$ if $f(\o)\in F(\o)$ for $\mu$-a.e. $\o \in \O$. We denote the set of all $\sE$-measurable selectors of $F$ by $\hS^0_\sE(F)$. Then, it follows from the standard measurable selection theorem that $\hS^0_\sE(F)\neq \emptyset$ whenever $F$ is measurable. Moreover, the following {\it Castaing Representation} is useful (cf. \cite{MK}):
{\it The set-valued function $F$ is measurable if and only if there exists a sequence $(f_n)_{n\in\hN}$ in $\hL^0_{\sE}(\O,\hR^d)$ such that $F(\o)=\cl\{f_n(\o)\colon n\in \mathbb{N}\}$ for $\mu$-a.e. $\o\in \O$}.

For $p\in[1,+\infty]$, let $\mathbb{L}_{\sE}^p(\O,\mathbb{R}^d)=\hL^p$ denote the space of all $f\in \hL_{\sE}^0(\O,\hR^d)$ such that $\|f\|_p:=(\int_{\O}|f(\o)|^p\mu(d\o))^{1/p}<\infty$ when $p\in [1,+\infty)$ and $\|f\|_\infty:=\inf\{\lambda>0\colon \hP\{|f|\leq \lambda\}=1\}<+\infty$ when $p=+\infty$. For a function
$F\in \scL^0_\sE(\O,\sC(\hR^d))$, we define $\hS^p_\sE(F):=\hS^0_\sE(F)\cap \mathbb{L}^p_\sE(\O,\hR^d)$ and say that $F$ is  {\it $p$-integrable} if $\hS^p_\sE(F)\neq \emptyset$. We denote $\sA^p_\sE(\O,\sC (\hR^d))$ to be the set of all 
$p$-integrable $\sC(\hR^d)$-valued functions.

An important concept in set-valued analysis is the so-called {\it decomposibility}. To be more precise, given a sub-$\sigma$-field $\sH\subseteq\sE$, a set $M\subseteq \mathbb{L}^0_\sE(\O,\hR^d)$ is said to be \emph{$\sH$-decomposable} if  $\1_Af+\1_{A^c}g\in M$ whenever $f,g\in M$ and $A\in\sH$; here, $\1_A(\o):=1$ if $\o\in A$ and $\1_A(\o)=0$ if $\o\in A^c$. For a set $M\subseteq \mathbb{L}^0_\sE(\O,\hR^d)$, we define the {\it $\sH$-decomposable hull} of $M$, denoted by $\dec_\sH(M) $, to be the smallest $\sH$-decomposable superset of $M$ in $\mathbb{L}^0_\sE(\O,\hR^d)$. When $M\subseteq \hL^p_{\sE}(\O,\hR^d)$ with $p\in [1,+\infty)$, we shall often consider the {\it closed $\sH$-decomposable hull} of $M$, denoted by  $\overline{\dec}_\sH(M):=\cl_{\hL^p}(\dec_\sH(M))$, where the closure is with respect to $\|\cdot\|_p$. The following theorem is  crucial in  our discussion below (cf. \cite{MK}).
 
\begin{thm}\label{thm:dec}
Let $M\in \sC(\mathbb{L}^p_\sE(\O,\hR^d))$, where $p\in [1,+\infty)$. Then, $M$ is $\sH$-decomposable if and only if there exists a unique $F\in \sA^p_\sH(\O,\sC (\hR^d))$ such that $M=\hS^p_\sH(F)$.
\end{thm}

Next, let us consider a complete filtered probability space $(\Omega,\mathcal{F},\mathbb{P},\mathbb{F}=(\mathcal{F}_t)_{t\in[0,T]})$, where $T>0$ is a constant. In this setting, a \emph{set-valued random variable} is meant to be a measurable function $\Xi\colon \O\to\sC(\hR^d)$ and a \emph{($\hF$-adapted) set-valued stochastic process} is defined as a collection $\Phi=(\Phi_t)_{t\in [0,T]}$, where $\Phi_t$ is a ($\mathcal{F}_t$-measurable) set-valued random variable for each $t\in[0,T]$, which can also be treated as a function $\Phi\colon [0,T]\times\O\to\sC(\hR^d)$. By a slight abuse of notation, we denote the $\hF$-progressive $\sigma$-field on $[0,T]\times\O$ by $\hF$. Accordingly, $\hL^0_{\hF}([0,T]\times\O,\hR^d)$ ($\scL^0_{\hF}([0,T]\times\O,\sC(\hR^d))$) denotes the set of all $\hF$-progressively measurable $\hR^d$-valued ($\sC(\hR^d)$-valued) processes. Given $\Phi\in \scL^0_{\hF}([0,T]\times\O,\sC(\hR^d))$, we denote $S^0_{\hF}(\Phi)$ to be the set of all $\hF$-progressively measurable selectors of $\Phi$, which is a nonempty set. For $p\in[1,+\infty)$, we define $\hS^p_{\hF}(\Phi):=\hS^0_{\hF}(\Phi)\cap \hL^p_{\hF}([0,T]\times\O,\hR^d)$ and
\[
\sA^p_{\hF}([0,T]\times \O, \sC(\hR^d))=\{\Phi\in \scL^0_{\hF}([0,T]\times \O,\sC(\hR^d))\mid \hS^p_\hF(\Phi)\neq\emptyset\}.
\]
For a vector-valued process $\phi\in \hL^p_{\hF}([0,T]\times\O,\hR^d)$ and $0\leq t\leq u\leq T$, let us define $J_{t,u}(\phi):=\int_t^u \phi_rdr$. Given $\Phi\in \sA^p_\hF([0,T]\times \O,\sC(\hR^d))$, the image of $\hS^p_{\hF}(\Phi)$ under $J_{t,u}$, i.e., the set
 \[
 J_{t,u}[\hS^p_{\hF}(\Phi)]:=\cb{\int_t^u \phi_rdr \colon \phi\in \hS^p_{\hF}(\Phi)}\subseteq \hL^p_{\cF_u}(\O,\hR^d)
 \]
 is called the \emph{functional set-valued integral} of $\Phi$ over $[t,u]$. By
 Theorem~\ref{thm:dec}, there exists a unique set-valued random variable $\int_t^u\Phi_rdr\in \sA^p_{\mathcal{F}_u}(\Omega,\sC (\hR^d))$ such that 
 \[
\hS^p_{\mathcal{F}_u}\of{\int_t^u\Phi_rdr}=\overline{\dec}_{\mathcal{F}_u}(J _{t,u} [\hS^p_{\hF}(\Phi)]);
 \]
 we call it the \emph{set-valued Lebesgue integral} of $\Phi$ over $[t,u]$.
 
\ms
\no{\bf Dynamic Set-Valued Risk Measures.} We recall some basic definitions related to dynamic set-valued risk measures; we refer the reader to \cite{FR13,FR2015} for a detailed account of the subject. Let us consider a filtered probability space $(\O, \cF, \hP, \hF=(\cF_t)_{t\in\hT})$, where we either have $\hT=[0,T]$ (continuous time) for some $T>0$ or $\hT$ is a finite subset of $[0,T]$ (discrete time). Let us fix $p\in [1,+\infty]$; we will use $p=2$ for the superhedging problem later. Let $t\in \hT$. 
For ease of notation, we write $\hL^p_{\cF_t}(A)$ for the set of all $\cF_t$-measurable $p$-integrable $A$-valued random vectors, where $A\subseteq \hR^d$ is nonempty. 
For $\xi,\eta\in \hL^p_{\cF_t}(\hR^d)$, we write $\eta\geq \xi$ whenever $\eta-\xi\in \hL^p_{\cF_t}(\hR^d_+)$. We also define the space of all \emph{upper sets} in $\hL^p_{\cF_t}(\hR^d)$ by
\[
\sP_+(\hL^p_{\cF_t}(\hR^d)):=\{M\subseteq \hL^p_{\cF_t}(\hR^d)\colon M=M+\hL^p_{\cF_t}(\hR^d_+)\},
\]
where we assume $M+\emptyset=\emptyset+M=\emptyset$ for each $M\subseteq \hL^p_{\cF_t}(\hR^d)$.

\begin{defn}\label{drm}
Let $t\in \hT$. For a set-valued functional $R_t\colon \hL^p_{\cF_T}(\hR^d)\rightarrow \sP_+(\hL^p_{\cF_t}(\hR^d))$, consider the following properties:

(i) \textbf{Monotone:} $Y\geq X$ implies $R_t(Y)\supseteq R_t(X)$ for every $X,Y\in \hL^p_{\cF_T}(\hR^d)$.

\ss
(ii) \textbf{Translative:} $R_t(X+\xi)=R_t(X)-\xi$ for every $X\in \hL^p_{\cF_T}(\hR^d)$ and $\xi\in\hL^p_{\cF_t}(\hR^d)$.

\ss
(iii) \textbf{Finite at zero:} $\emptyset\neq R_t(0)\neq \hL^p_{\cF_t}(\hR^d)$.

\ss
(iv) \textbf{Normalized:} $R_t(X)=R_t(X)+R_t(0)$ for every $X\in \hL^p_{\cF_T}(\hR^d)$.

\ss
(v) \textbf{Conditionally convex:} $R_t(\lambda X+(1-\lambda)Y)\supseteq \lambda R_t(X)+(1-\lambda)R_t(Y)$ for every $X,Y\in \hL^p_{\cF_T}(\hR^d)$ and $\lambda\in \hL^{\infty}_{\cF_t}([0,1])$.

\ss
(vi) \textbf{Conditionally positively homogeneous:} $R_t(\lambda X)=\lambda R_t(X)$ for every $X\in \hL^p_{\cF_t}(\hR^d)$ and $\lambda\in \hL^{\infty}_{\cF_t}(\hR_{++})$.

\ss 
(vii) \textbf{Closed:} $\gr R_t:=\{(X,\xi)\in\hL^p_{\cF_T}(\hR^d)\times\hL^p_{\cF_t}(\hR^d)\colon \xi\in R_t(X)\}$ is a closed set in the product topology.

\ss 
(viii) \textbf{Decomposable:} $R_t(\1_A  X+\1_{A^c}Y)=\1_A R_t(X)+\1_{A^c}R_t(Y)$ for every $X,Y\in \hL^p_{\cF_T}(\hR^d)$ and $A\in \cF_t$.

The functional $R_t$ is called a \textbf{conditional set-valued risk measure at time $t$} if it satisfies (i), (ii), (iii). In this case, $R_t$ is called \textbf{conditionally coherent} if it also satisfies (v), (vi).

A family $(R_t)_{t\in\hT}$ is called a \textbf{(conditionally coherent) dynamic set-valued risk measure} if $R_t$ is a (conditionally coherent) conditional set-valued risk measure at time $t$ for each $t\in\hT$.
\qed
\end{defn}

The properties in Definition~\ref{drm} are the multi-dimensional and set-valued analogs of the properties of conditional risk measures for univariate positions. A noticeable aspect is that a larger set indicates lower risk as there are more portfolios that can be used for risk compensation. In particular, monotonicity indicates that a financial position with larger gains is less risky and conditional convexity reflects the principle that diversification reduces risk. Translativity and positive homogeneity read the same as their univariate counterparts with obvious interpretations. Finiteness at zero ensures that the zero portfolio has at least one risk compensating portfolio and not every portfolio can be used for this purpose. Finally, closedness is a lower semicontinuity property that is essential in obtaining dual representations for convex and coherent risk measures.

\begin{rem}\label{rem:drm}
	{\rm Let $(R_t)_{t\in\hT}$ be a conditionally convex dynamic set-valued risk measure. Let $t\in\hT$ and $X\in \hL^p_{\cF_T}(\hR^d)$. It can be checked that $R_t(X)$ is an $\cF_t$-decomposable convex set in $\hL^p_{\cF_t}(\hR^d)$. In particular, $R_t$ maps into the space of all convex upper sets given by
		\[
		\mathscr{F}_+(\hL^p_{\cF_t}(\hR^d)):=\{M\subseteq \hL^p_{\cF_t}(\hR^d)\colon M=\co(M+\hL^p_{\cF_t}(\hR^d_+))\}.
		\]
		Moreover, by Theorem~\ref{thm:dec}, there exists a set-valued random variable $\tilde{R}_t(X)\in\sA^p_{\cF_t}(\O,\sG(\hR^d))$ such that
		\[
		\cl_{\hL^p}(R_t(X))=\hS^p_{\cF_t}(\tilde{R}_t(X)).
		\] 
		This way, every $X\in \hL^p_{\cF_T}(\hR^d)$ gives rise to an $\hF$-adapted set-valued stochastic process $(\tilde{R}_t(X))_{t\in\hT}$ that determines $(\cl_{\hL^p}(R_t(X)))_{t\in\hT}$.
		\qed}
\end{rem}

Given a conditional risk measure $R_t$ at time $t\in\hT$, we define its \emph{acceptance set} by
\[
A_{t}:=\{X\in\hL^p_{\cF_T}(\hR^d)\colon 0\in R_t(X)\},
\] 
which gives the set of all $d$-dimensional financial positions at time $t$ that are considered acceptable. Then, $R_t$ can be recovered from its acceptance set via
\[
R_t(X)=\{\xi\in\hL^p_{\cF_t}(\hR^d)\colon X+\xi\in A_t\}
\]
for each $X\in\hL^p_{\cF_t}(\hR^d)$; hence, each $\xi\in R_t(X)$ can be seen as a portfolio at time $t$ that can be used to compensate the risk of $X$. If $u\in\hT$ is such that $t\leq u$, then we denote $R_{t,u}$ to be the restriction of $R_t$ on the subspace $\hL^p_{\cF_u}(\hR^d)$, also called a \emph{stepped risk measure}. In this case, the corresponding \emph{stepped acceptance set} is defined by $A_{t,u}:=\{\eta \in\hL^p_{\cF_u}(\hR^d)\colon 0\in R_{t}(\eta)\}$ since $R_{t,u}(\eta)=R_t(\eta)$ for every $\eta\in\hL^p_{\cF_u}(\hR^d)$.

A dynamic risk measure $(R_t)_{t\in \hT}$ is called \emph{multi-portfolio time-consistent} if, for every $t<u$ in $\hT$, $X\in \hL^p_{\cF_T}(\hR^d)$, and $M\subseteq\hL^p_{\cF_T}(\hR^d)$, the following implication holds:
\[
R_u(X)\subseteq \bigcup_{Y\in M}R_u(Y)\implies R_t(X)\subseteq \bigcup_{Y\in M}R_t(Y)=:R_t[M].
\]
Multi-portfolio time-consistency is the proper extension of time-consistency for set-valued risk measures and it can be characterized by a recursive property as recalled next.

\begin{thm}
\label{ConRm}
\cite[Theorem~2.8]{FR2015} For a normalized dynamic set-valued risk measure $(R_t)_{t\in \hT}$ the following are equivalent:

\ss
(i) $(R_t)_{t\in\hT}$ is multi-portfolio time consistent.

\ss
(ii) $R_t(X)=R_t[-R_u(X)]$ for every $t<u$ in $\hT$ and $X\in \hL^p_{\cF_T}(\hR^d)$.

\ss
(iii) $A_t=A_{t,u}+A_u$ for every $t<u$ in $\hT$.
\end{thm}

 \ms
 \no{\bf The Superhedging Problem in Discrete Time.} Before formulating our \emph{superhedging problem} in continuous time, we shall first recall its discrete-time counterpart as discussed in \cite {FR13} and \cite{lohnerudloff}. For simplicity, let us assume that $T\in\hN$ and take $\hT=\{0,\ldots,T\}$ in the setting of the previous subsection. For each $t\in\hT$, let $\hat{K}_t\in \scL_{\cF_t}^0(\O,\sG(\hR^d))$ be a random convex cone such that $\hR^d_+\subseteq \hat{K}_t(\o)$ for $\hP$-a.e. $\o\in\O$. We call $\hat{K}_t$ the \emph{solvency cone} at time $t\in\hT$; each element of $\hS^p_{\cF_t}(\hat{K}_t)$ is a portfolio vector at time $t$ that can be exchanged into a portfolio vector with long positions in all assets, where $p\geq 1$. An $\hR^d$-valued process $(\hat{V}_t)_{t\in\hT}$ is said to be a \emph{self-financing portfolio process} if $\hat{V}_t-\hat{V}_{t-1}\in \hS^p_{\cF_t}(-\hat{K}_t)$ for each $t\in\hT$, where we set $\hat{V}_{-1}:=0$. Then, the set of all terminal values of self-financing and $p$-integrable portfolio processes with zero value at time $t\in\hT$ is given by $C_{t,T}:=\sum_{r=t}^T \hS^p_{\cF_r}(-\hat{K}_r)$. For a financial position $X\in\hL^p_{\cF_T}(\hR^d)$, the set of all \emph{superhedging portfolios} of it at time $t\in\hT$ is given by
\bea\label{SHP}
SHP_t(X):=\{\xi\in\hL^p_{\cF_t}(\hR^d)\colon X\in \xi+C_{t,T}\}.
\eea 
By \cite[Corollary~5.2]{FR13}, under the robust no-arbitrage condition (see \cite[Section~5.1]{FR13} for the precise formulation), the family $(R_t)_{t\in\hT}$ defined by $R_t(X):=SHP_t(-X)$, for each $X\in\hL^p_{\cF_T}(\hR^d)$ and $t\in\hT$, is a normalized, closed, conditionally coherent dynamic set-valued risk measure that is also multi-portfolio time-consistent.

\section{Solvency Cone and Consistent Prices}\label{sec:market}
\setcounter{equation}{0}

In this section, we introduce the continuous-time market model on which we will study the superhedging problem. Specifically, using the notion of a \emph{solvency cone} (cf. \cite{KandS,Sch}), we will consider a generalized Black-Scholes-type model that is a multi-asset version of the model in \cite{cvi}.

 \ms
\no{\bf The Continuous-Time Model.}
Let us fix a probability space $(\O,\cF,\hP)$ on which there exists an $m$-dimensional standard Brownian motion $(W_t)_{t\in[0,T]}$, where $m\in \hN$ and $T>0$ is a finite deterministic horizon. Let $\hF=(\cF_t)_{t\in[0,T]}$ be the standard filtration of $(W_t)_{t\in[0,T]}$ augmented by the $\hP$-null sets of $\cF$. In this setting, we consider a financial market that consists of $d\in \hN$ assets, where asset $1$ is fixed as a \emph{num\'{e}raire}. The prices of the assets, quoted in terms of asset $1$, are described by a $d$-dimensional, strictly positive, $\hF$-adapted process $(S_t)_{t\in [0,T]}$ with $S_t=(S^1_t, \ldots, S^d_t)$ for each $t\in [0,T]$. We assume that $(S_t)_{t\in[0,T]}$ has the following generalized Black-Scholes dynamics with random coefficients:
\bea
\label{BS}
\left\{\ba{lll}
\dis dS^1_t=r_tS^1_tdt, \qq\qq S^1_0=1, \ms \\
\dis dS^i_t=b^i_t S^i_t dt+\sum_{\ell=1}^m\si^{i\ell}_t S^i_t dW^\ell_t,~~~S^i_0=s_0^i>0, \q i\in [d]\setminus\{1\},
\ea\right.
\eea
where $[d]:=\{1,\ldots,d\}$; $(b_t)_{t\in [0,T]}=(r_t,b^2_t,\ldots,b^d_t)_{t\in[0,T]}$ and $(\si_t)_{t\in [0,T]}$ are 
$\hF$-progressively measurable bounded processes with values in $\hR^d$ and $\hR^{d\times m}$, respectively; $\si^{1\ell}\equiv 0$ for each $\ell\in [m]$. In matrix notation, we may rewrite \eqref{BS} as
\bea
\label{BScompact}
dS_t=\diag(S_t)\of{b_t dt+\si_t dW_t},~~~S_0=s_0:=(1,s_0^2,\ldots,s_0^d),
\eea
where, for $x\in\hR^d$, $\diag(x)$ denotes the $d\times d$-matrix whose $i^{\text{th}}$ diagonal entry is $x_i$ for each $i\in[d]$ and all other entries are zero. By the general well-posedness results for stochastic differential equations, the boundedness assumptions on the coefficients guarantee that the above dynamics determine $(S_t)_{t\in[0,T]}$ uniquely as an $\hF$-adapted continuous process with $\hE[\sup_{t\in[0,T]}|S_t|^2]<+\infty$; see \cite[Theorem~5.2.9]{karatzas}, for instance. We assume that $m=d-1$ for simplicity.

 \ms
\no{\bf  Transactions Costs and the Solvency Cone.} We further assume that asset $1$ also plays the role of {\it bank account} to be used for the accounting of transaction fees, which we now describe. Let $i,j\in [d]$ and suppose that the market allows transferring funds from asset $i$ to asset $j$. In case of such a transfer, a transaction fee has to be deducted from the bank account according to a given \emph{deterministic} proportion $\mu^{ij}\in[0,1)$. We assume that $\mu^{ii}=0$ and, for every $k\in[d]$,
\bea 
\label{triangle}
(1+\mu^{ij})\leq (1+\mu^{ik})(1+\mu^{kj}),
\eea 
i.e., an indirect transfer through a third asset never reduces the incurred transaction cost.

In this market model, we describe trading strategies via cumulative fund transfers between assets. Let $i,j\in[d]$. We denote by $L^{ij}_t$ the net cumulative amount of funds, quoted in asset $1$, that is transferred from asset $i$ to asset $j$ during $[0,t]$, where $t\in[0,T]$. 
We assume that $L^{ij}=(L^{ij}_t)_{t\in[0,T]}$ is an $\hF$-adapted, c\`adl\`ag (right-continuous and left-limited), nondecreasing process such that $L^{ij}_0=0$. Naturally, we assume that $L^{ii}\equiv 0$.
Then, given an initial position $V_0=v:=(v^1,\ldots,v^d)\in\hR^d$, a $d$-dimensional process $(V_t)_{t\in[0,T]}$, quoted in asset $1$, is called {\it a self-financing portfolio process} if it has the following dynamics for some trading strategy $(L^{ij})_{i,j\in[d]}$:
\bea
\label{V1}
\left\{\ba{lll}
\dis dV^1_t=V^1_t r_t dt-\sum_{j=2}^d(1+\mu^{1j})dL^{1j}_t+\sum_{j=2}^d(1-\mu^{j1})dL^{j1}_t-\sum_{i=2}^d\sum_{j=2}^d \mu^{ij}dL^{ij}_t, \ms \\
\dis dV^i_t=   V^i_t b^i_t dt+\sum_{\ell=1}^m  V^i_t\si^{i\ell}_t  dW^\ell_t+\sum_{j=1}^d (dL^{ji}_t-dL^{ij}_t), \q i\in [d]\setminus\{1\}.
\ea\right.
\eea
We remark that, in (\ref{V1}), the terms involving the funds $L^{j1}$ are transferred from assets $j$, therefore they are subject to the transaction fees.

In light of the notion of {\it solvency cone} by \cite{Kcurrency, KandS, Sch} and a 2-dimensional special case studied in \cite{AlMa}, we now construct a  deterministic solvency cone in  $\hR^d$ based on the portfolio dynamics \reff{V1}. To begin with, let us define the following \emph{exchange matrix}
in accordance with (\ref{V1}) (in the spirit of a \emph{bid-ask matrix} as in \cite{Sch} but with some negative entries in our setting), noting that $\m^{ii}=0$:
\beaa
\Pi=(\pi^{ij})_{i,j\in[d]}:=\begin{pmatrix}
1&1+\mu^{12}& 1+\mu^{13}&\cdots & 1+\mu^{1d}\\
-(1-\mu^{21})& 0& \m^{23}&\cdots& \m^{2d}\\
-(1-\mu^{31})&\m^{32} &0&\cdots &\m^{3d}\\
\vdots  & \vdots  & \vdots & \ddots &\vdots \\
-(1-\mu^{d1})&\m^{d2}&\m^{d3}&\cdots&0
\end{pmatrix}.
\eeaa 
Next, given matrices $A =(a^{ij})_{i,j\in[d]}, B =(b^{ij})_{i,j\in[d]}\in \hR^{d\times d}$, we define their Frobenius inner product by
\[
\ip{A,B}:=\tr(A^{\mathsf{T}}B)=\sum_{i=1}^d \sum_{j=1}^d a^{ij}b^{ij}.
\]
For each $i\in[d]$, we denote by $e_i$ the standard $i^{\text{th}}$ unit (column) vector in $\hR^d$, we also define $E_i:=[e_i, \ldots, e_i]$ to be the $d\times d$ matrix whose entries in the $i^{\text{th}}$ row are $1$ and all other entries are $0$. Let us denote $\hM^d_+\subset \hR^{d\times d}_+$ to be the convex cone of all $d\times d$ matrices $A=(a^{ij})_{i,j\in [d]}$ with $a^{ij}\ge 0$ and $a^{ii}=0$ for every $i,j\in [d]$. We define the convex cone
\bea
\label{KPi}
K(\Pi):=\{x \in\hR^d\colon x^1=\ip{A,\Pi}, ~x^i=\langle A-A^{\mathsf{T}}, E_i\rangle \ \forall i\in [d]\setminus\{1\},~ A\in\hM^{d}_+ \},
\eea
and call it the {\it solvency cone} associated to the bid-ask matrix $\Pi$.

The following  result regarding the solvency cone will be  useful for our discussion.

\begin{prop}\label{prop:cone-upper}
\label{K=M}
$K(\Pi)=K(\Pi)+\hR^d_+$. 
\end{prop}

{\it Proof.} Clearly, $K(\Pi)\subseteq K(\Pi)+\hR^d_+$, since $0\in\hR^d_+$. To show that $K(\Pi)+\hR^d_+\subseteq K(\Pi)$, let $x:=y+\a$ for some $y\in K(\Pi)$ and  $\a\in\hR^d_+$. By (\ref{KPi}), there exists some $A=(a^{ij})_{i,j\in[d]}\in\hM^{d}_+$ such that $y^1=\ip{A,\Pi}$ and $y^i=\langle A-A^{\mathsf{T}},E_i\rangle$ for each $i\in [d]\setminus\{1\}$. Hence,
\beaa 
\left\{\ba{lll}
\dis x^1-\sum_{j=2}^da^{1j}(1+\mu^{1j})+\sum_{j=2}^da^{j1}(1-\mu^{j1})-\sum_{i=2}^d\sum_{j=2}^d a^{ij}\mu^{ij}=\alpha^1,\ms \\
\dis x^i+\sum_{j=1}^d (a^{ji}-a^{ij})=\alpha^i ,~~~i\in[d]\setminus \{1\}.
\ea\right.
\eeaa 
Note that if we can show that $\a\in K(\Pi)$ as well, 
then we must have $x=y+\a\in K(\Pi)$, since $K(\Pi)$ is a convex cone, proving the claim. It now suffices to show that the following system of algebraic equations has a solution $B\in\hM^{d}_+$:
\bea
\label{eqB}
\left\{\ba{lll}
\dis\ip{B,\Pi}=
\sum_{j=2}^db^{1j}(1+\mu^{1j})-\sum_{j=2}^db^{j1} (1-\mu^{j1})+\sum_{i=2}^d\sum_{j=2}^d b^{ij}\mu^{ij}=\a^1,\ms \\
\dis \langle B-B^{\mathsf{T}}, E_i \rangle =\sum_{j=1}^d (b^{ij}-b^{ji})=\alpha^i,~~~ i\in[d]\setminus\{1\}.
\ea\right.
\eea
To this end, since all diagonal entries of a matrix in $\hM^d_+$ are zero, we can embed the cone $\hM^d_+$ into $\hR^{d(d-1)}_+$ by the mapping $\hM^d_+\ni B\mapsto \bi_B \in \hR^{d(d-1)}_+$ defined by
\[
\bi_B:=\big(b^{12},\ldots, b^{1d}; b^{21},\ldots,b^{d1};b^{23},b^{24},\ldots,b^{d(d-2)}b^{d(d-1)}\big)^{\mathsf{T}} \in \hR^{d(d-1)}_+.
\]
We can then rewrite the linear system \eqref{eqB} in the usual form
\bea
\label{eqB1}
C(\Pi) \bi_B=\a,
\eea
where $C(\Pi)$ is the coefficient matrix of \eqref{eqB}, that is,
{\footnotesize
	\[
	C(\Pi)=\begin{pmatrix}
		1+\mu^{12} & \cdots & 1+\mu^{1d}&\negthinspace-(1-\mu^{21})&\cdots& \negthinspace -(1-\mu^{d1})&\mu^{23}&\mu^{24}&\cdots&\mu^{d(d-2)}&\mu^{d(d-1)}\\
		-1& \cdots& 0 &1&\cdots&0&1 &1&\cdots & 0 &0\\
		\vdots  & \ddots  & \vdots & \vdots &\ddots&\vdots&\vdots&\vdots& &\vdots & \vdots\\
		0 & \cdots & -1&0&    \cdots &1 &0&0&\cdots &-1 &-1
	\end{pmatrix}.
	\]
}
	
	Now let us look at the following dual algebraic problem: find $y\in\hR^d$ such that 
	\bea
	\label{prob1}
	C(\Pi)^{\mathsf{T}}y\in\hR^{d(d-1)}_+, \qq \a^{\mathsf{T}}y<0.
	\eea
	Writing $y=(y^1, \ldots, y^d)^{\mathsf{T}}$, we see that
	$C(\Pi)^{\mathsf{T}}y\in\hR^{d(d-1)}_+$ implies that $(1+\mu^{12})y^1-y^2\geq 0$ and $-(1-\mu^{21})y^1+y^2\geq 0$. Thus, $(\mu^{12}+\mu^{21})y^1\geq 0$ and therefore $y^1\geq 0$, as $\mu^{12}+\mu^{21}\ge0$. Furthermore, we also have  $-(1-\mu^{j1})y^1+y^j\geq 0$, whence $y^j\geq (1-\mu^{j1})y^1 \geq 0$ for every $j\in [d]\setminus\{1\}$. In other words, $y\in\hR^d_+$, and consequently, we have $\alpha^{\mathsf{T}}y \geq 0$, since $\a\in\hR^d_+$ as well. Therefore, the problem (\ref{prob1}) does not have a solution. Now, by Farkas' lemma (cf., e.g., \cite[p.~201]{rockafellar}), the linear algebraic equation
	\eqref{eqB1} must have a solution $\bi_B\in \hR^{d(d-1)}_+$, or equivalently, the equation (\ref{eqB}) has a solution $B\in\hM^d_+$, proving the claim $K(\Pi)+\hR^d_+\subseteq K(\Pi)$, whence the proposition. 
\qed

 \ms
\no{\bf  Instantaneous Trading and Self-Financing Portfolios.} In what follows, we will restrict ourselves to trading strategies that occur at an instantaneous rate in continuous time. For this purpose, let $\hU_{\hF}$ be the set of all $(L^{ij})_{i,j\in[d]}$ such that, for each $i,j\in[d]$, $L^{ij}_0=0$ and
\[
dL^{ij}_t=\theta^{ij}_tdt
\]
for some process $\theta^{ij}\in \hL^2_{\hF}([0,T]\times\O,\hR_+)$, with $\theta^{ii}\equiv 0$. The next theorem characterizes self-financing portfolio processes in this setting. To that end, for a $d$-dimensional process $(V_t)_{t\in[0,T]}$, quoted in asset $1$, we denote by $(\hat{V}_t)_{t\in[0,T]}$ the corresponding process quoted in physical units, i.e.,
\bea\label{eq:hat}
\hat{V}_t=\of{\frac{V_t^1}{S_t^1},\ldots,\frac{V_t^d}{S_t^d}}^{\mathsf{T}},\quad t\in[0,T].
\eea
We also define a set-valued process $\hat{K}=(\hat{K}_t)_{t\in[0,T]}$ via
\bea\label{eq:Khat}
\hat{K}_t(\o):=\cb{\of{\frac{x^1}{S_t^1(\o)},\ldots,\frac{x^d}{S_t^d(\o)}}\colon x\in K(\Pi)},\quad (t,\o)\in[0,T]\times\O,
\eea
and the corresponding $\hL^2$-space of vector-valued processes in physical units via
\beaa
\hat{\hS}^2_{\hF}(\hat{K}):=\cb{\hat{k}\in \hS^0_{\hF}(\hat{K})\colon \diag(S)\hat{k}\in \hL^2_{\hF}([0,T]\times\O,\hR^d)}.
\eeaa

\begin{rem}
{\rm As a consequence of Itô's formula, we have $(\frac{1}{S^1},\ldots,\frac{1}{S^d})\in\hL^2_{\hF}([0,T]\times\O,\hR^d)$. In particular, the inclusion $\hat{\hS}^2_{\hF}(\hat{K})\subseteq \hS^1_{\hF}(\hat{K})$ holds.
	\qed}
\end{rem}

\begin{thm}
	\label{Vhat}
Let $(V_t)_{t\in[0,T]}$ be a $d$-dimensional process with $v:=V_0$ and $\hat{v}:=\hat{V}_0$. Then, the following are equivalent:\\
		(i) $(V_t)_{t\in[0,T]}$ is a self-financing portfolio process, i.e., there exists $(L^{ij})_{i,j\in[d]}\in\hU_{\hF}$ such that \eqref{V1} holds.\\
		(ii) There exists $\Th\in \hS^2_{\hF}(\hM^d_+)$ such that
		\bea
		\label{V3}
		\left\{\ba{lll}
		\dis dV^1_t= (V^1_tr_t-\ip{\Th_t,\Pi})dt,\ms \\
		\dis dV^i_t=V^i_t b^i_tdt +\sum_{\ell=1}^m V_t^i \si^{i\ell}_t dW^{\ell}_t-\langle \Th_t-\Th_t^{\mathsf{T}},E_i\rangle dt,~~~ i\in[d]\setminus\{1\}.
		\ea\right.
		\eea
		(iii) There exists $k\in \hS^2_{\hF}(K(\Pi))$ such that
		\bea
		\label{Vdyn2}
		\left\{\ba{lll}
		\dis dV^1_t=(V^1_t r_t-k_t^1)dt,\ms \\
		\dis dV^i_t=V^i_t b^i_tdt +\sum_{\ell=1}^m V_t^i \si^{i\ell}_t dW^{\ell}_t-k_t^i dt,~~~ i\in[d]\setminus\{1\}.
		\ea\right.
		\eea
		(iv) $(V_t)_{t\in[0,T]}$ satisfies, for each $t\in[0,T]$,
		\bea
		\label{Vinclus-func}
		V_t\in v+\int_0^t \diag(V_s)b_s ds  - J_{0,t}[\hS^2_{\hF}(K(\Pi))]+\int_0^t \diag(V_s)\si_sdW_s. 
		\eea
		(v) $(\hat{V}_t)_{t\in[0,T]}$ satisfies
		\bea
		\label{VhatinK1}
		\hat{V_t} \in \hat{v} +J_{0,t}[\hat{\hS}^2_{\hF}(-\hat{K})],~~~~~~ t\in [0,T].
		\eea
		In this case, $(\hat{V}_t)_{t\in[0,T]}$ also satisfies
		\[
		\hat{V}_t\in \hat{v}+\int_0^t (-\hat{K}_s)ds\q a.s.~~~~~~ t\in [0,T].
		\]
\end{thm}

 {\it Proof.}  (i) $\Leftrightarrow$ (ii): Suppose that there exists some $(L^{ij})_{i,j\in[d]}\in\hU_{\hF}$ such that \eqref{V1} holds. Then, for each $i,j\in[d]$, we can find $\theta^{ij}\in\hL^2_{\hF}([0,T]\times\O,\hR_+)$ such that $dL_t^{ij}=\theta_t^{ij}dt$, where $\theta^{ii}\equiv 0$. Hence, \eqref{V1} can be rewritten as
 \bea
 \label{V2}
 \left\{\ba{lll}
 \dis  dV^1_t=V^1_t r_t dt\neg-\neg\sum_{j=2}^d(1+\mu^{1j})\theta^{1j}_t dt +\sum_{j=2}^d(1-\mu^{j1})\theta^{j1}_tdt\neg-\neg\sum_{i=2}^d\sum_{j=2}^d \mu^{ij}\theta^{ij}_tdt,\ms \\
 \dis  dV^i_t=   V^i_t b^i_t dt+\sum_{\ell=1}^m  V^i_t\si^{i\ell}_t  dW^\ell_t+\sum_{j=1}^d (\theta^{ji}_t -\theta^{ij}_t)dt, \q i\in [d]\setminus\{1\}.
 \ea\right.
 \eea
 Defining $\Theta_t=(\theta^{ij}_t)_{i,j\in[d]}$ for each $i,j\in[d]$ and $t\in[0,T]$, and recalling \eqref{eqB}, we observe that $\Theta\in \hS^2_{\hF}(\hM^d_+)$ and \eqref{V2} can be reformulated as \eqref{V3}. The reverse implication follows similarly.
 
 (ii) $\Leftrightarrow$ (iii): Suppose that there exists $\Theta\in \hS^2_{\hF}(\hM^d_+)$ such that \eqref{V3} (equivalently, \eqref{V2}) holds. For each $t\in [0,T]$, let us define $k_t:=(k_t^1,\ldots,k_t^d)^{\mathsf{T}}$ by
 \[
 k_t^1:=\ip{\Theta_t,\Pi},\quad k_t^i:=\langle \Theta-\Theta^{\mathsf{T}},E_i\rangle,\ i\in [d]\setminus\{1\}.
 \]
 By the definition of the solvency cone in \eqref{KPi}, we have $k=(k_t)_{t\in[0,T]}\in \hS^2_{\hF}(K(\Pi))$. Moreover, \eqref{Vdyn2} follows directly by \eqref{V2}. The reverse implication follows similarly.
 
 (iii) $\Leftrightarrow$ (iv): Suppose that there exists $k\in \hS^2_{\hF}(K(\Pi))$ such that \eqref{Vdyn2} holds. Recalling the notation in \eqref{BScompact}, we may rewrite \eqref{Vdyn2} as
 \[
 dV_t=(\diag(V_t)b_t-k_t)dt+\diag(V_t)\sigma_t dW_t.
 \]
 Hence, for each $t\in[0,T]$, we have
 \begin{equation*}
 	\begin{aligned}
 		V_t&=v+\int_0^t \diag(V_s)b_sds -\int_0^t k_sds +\int_0^t \diag(V_t)\sigma_t dW_t\\
 		& \in v+ \int_0^t \diag(V_s)b_sds-J_{0,t}[S^2_{\hF}(K(\Pi))]+\int_0^t \diag(V_s)\sigma_s dW_s
 	\end{aligned}
 \end{equation*} 
 so that \eqref{Vinclus-func} holds. The reverse implication follows similarly.
 
 (iii) $\Rightarrow$ (v): Suppose that there exists $k\in \hS^2_{\hF}(K(\Pi))$ such that \eqref{Vdyn2} holds. By It\^{o}'s formula and using the price dynamics in \eqref{BS}, we obtain that $(\hat{V}_t)_{t\in[0,T]}$ has the dynamics  
 \bea
 \label{dhatV}
 \left\{\ba{lll}
 \dis  d\hat V^1_t=d\Big(\frac{V^1_t}{S^1_t}\Big)=\frac{1}{S^1_t} (r_tV^1_t- k^1_t)dt-V^1_t  \Big(\frac{1}{S^1_t}\Big)r_tdt =-\frac{k_t^1}{S^1_t} dt,\ms \\
 \dis d\hat V^i_t=d\Big(\frac{V^i_t}{S^i_t}\Big)=\frac{1}{S^i_t} dV^i_t+V^i_t d\Big(\frac{1}{S^i_t}\Big)-\frac{\sum_{\ell=1}^m (\sigma^{i\ell}_t)^2V^i_t}{S^i_t} dt 
 =-\frac{k_t^i}{S^i_t}  dt,
 \ea\right.
 \eea
 for every $i\in[d]\setminus\{1\}$.
 
 Using the notation in \eqref{eq:hat}, we define the process $\hat{k}$ corresponding to $k$. Then, we have $\hat{k}\in \hat{\hS}^2_{\hF}(\hat{K})$ and \eqref{dhatV} can be rewritten as
 \bea\label{khat}
 d\hat{V}_t=-\hat{k}_tdt,
 \eea 
 which implies that \eqref{VhatinK1} holds. 
 
 (v) $\Rightarrow$ (iii): Suppose that $(\hat{V}_t)_{t\in[0,T]}$ satisfies \eqref{VhatinK1}. Hence, there exists $\hat{k}\in \hat{\hS}^2_{\hF}(\hat{K})$ such that \eqref{khat} holds. Then, by It\^{o}'s formula and \eqref{BS}, we obtain
 \[
 \left\{\ba{lll}
 \dis dV^1_t=d(\hat{V}^1_t S^1_t)=\hat{V}_t^1 dS_t^1+S_t^1d\hat{V}_t^1=V_t^1r_tdt-k_t^1dt,\ms \\
 \dis dV^i_t=d(\hat{V}^i_t S^i_t)=\hat{V}_t^i dS_t^i+S_t^id\hat{V}_t^i+(dS_t^i)(d\hat{V}_t^i)=V^i_t b^i_tdt +\sum_{\ell=1}^m V_t^i \si^{i\ell}_t dW^{\ell}_t-k_t^i dt,
 \ea\right.
 \]
 for every $i\in[d]\setminus\{1\}$. Hence, $(V_t)_{t\in[0,T]}$ satisfies \eqref{Vdyn2}.
\qed 

\ms
\no{\bf The Dual of Solvency Cone and Consistent Price Processes.} In the rest of the section, we shall study the set-valued process of the dual cones of $\hat K=(\hat K_t)_{t\in[0,T]}$, and define the so-called consistent pricing processes. To begin with, by Proposition~\ref{K=M} and \eqref{eq:Khat}, for each $(t,\o)\in[0,T]\times\O$, the cone $\hat{K}_t(\o)$ (in physical units) consists of all $x\in \hR^d$ for which
\bea
\label{hatK}
x^1_t=\frac{1}{S^1_t(\o)}\ip{A,\Pi}, \q
x^i_t=\frac{1}{S^i_t(\o)}\langle A-A^{\mathsf{T}},E_i\rangle , \ i\in[d]\setminus\{1\}.
\eea
holds for some $A\in\hM^d_+$. We define the (positive) dual cone of $\hat{K}_t(\o)$ by
\bea
\label{hatK*}
\hat{K}^+_t(\o)
:=\{z\in \hR^d\colon  z^{\mathsf{T}}x\geq 0\text{ for every }x \in \hat{K}_t(\o)\}. 
\eea
The following proposition shows that the dual cones are in line with the {\it auxiliary martingales} introduced for the two-dimensional continuous-time case in \cite{cvi}.

\begin{prop}\label{prop:dualcone}
Let $t\in[0,T]$, $\o\in\O$, and $z\in\hR^d$. Then $z\in \hat{K}^+_t(\o)$ if and only if 
\beaa
(1-\mu^{i1})\frac{z^1}{S^1_t(\o)}\leq \frac{z^i}{S^i_t(\o)}\leq (1+\mu^{1i})\frac{z^1}{S^1_t(\o)}, \q
 \frac{z^j}{S^j_t(\o)}- \frac{z^i}{S^i_t(\o)}\leq \mu^{ij}\frac{z^1}{S^1_t(\o)}, \ i,j\in[d]\setminus\{1\}.
 \eeaa
 \end{prop}
 
 {\it Proof.} By \eqref{hatK} and \eqref{hatK*}, we have $z\in \hat{K}_t^+(\o)$ if and only if the following inequality holds for every $A=(a^{ij})_{i,j\in[d]}\in\hM^d_+$:
 \[
 \frac{z^1}{S_t^1(\o)}\ip{A,\Pi}+\sum_{i=1}^d \frac{z^i}{S_t^i(\o)}\langle A-A^{\mathsf{T}},E_i\rangle \geq 0. 
 \]
 Recalling \eqref{eqB} and reorganizing the terms, we may rewrite this inequality as
 \bea\label{eq:dual}
 \sum_{j=2}^d c^{1j}(z)a^{1j}+\sum_{i=2}^d c^{i1}(z)a^{i1}+\sum_{i=2}^d\sum_{j=2}^d c^{ij}(z)a^{ij}\geq0,
 \eea
 where
 \[
 \left\{\ba{lll}
 \dis c^{1j}(z):=(1+\mu^{1j})\frac{z^1}{S^1_t(\o)}-\frac{z^j}{S^j_t(\o)}, \q c^{i1}(z):=-(1-\mu^{i1})\frac{z^1}{S^1_t(\o)}+\frac{z^i}{S^i_t(\o)},\q  \ms\\
 \dis c^{ij}(z):=\mu^{ij}\frac{z^1}{S^1_t(\o)}+\frac{z^i}{S^i_t(\o)}-\frac{z^j}{S^j_t(\o)}\geq 0, \q i,j \in [d]\setminus\{1\}.
 \ea\right.
 \]
 Hence, $z\in \hat{K}_t^+(\o)$ if and only if \eqref{eq:dual} holds for every $a^{1j},a^{i1},a^{ij}\geq 0$ and $i,j\in[d]\setminus\{0\}$, or equivalently, the corresponding coefficients are nonnegative, i.e., $c^{1j}(z)\geq 0$, $c^{i1}(z)\geq 0$, $c^{ij}(z)\geq 0$ for every $i,j\in[d]\setminus\{1\}$. Therefore, the result follows.
\qed

The preceding proposition implies that the nonzero elements of the dual of the solvency cone are always strictly positive.
	
\begin{cor}
\label{cor:positive}
It holds $\hP\{\hat{K}_t^+\setminus\{0\}\subseteq \hR^d_{++}\text{ for every } t\in[0,T]\}=1$.
\end{cor}

{\it Proof}. The generalized Black-Scholes dynamics of $S$ in \eqref{BS} guarantee that, for some $\O_0\in\cF$ with $\hP(\O_0)=1$, we have $S^i_t(\o)>0$ for every $t\in[0,T]$, $i\in[d]$, and $\o\in\O_0$. Let us fix $t\in[0,T]$, $\o\in\O_0$ and let $z\in \hat{K}_t^+(\o)\setminus\{0\}$. Suppose that $z^1=0$. Then, the first inequality in Proposition~\ref{prop:dualcone} yields $z^i=0$ for every $i\in [d]\setminus\{1\}$ so that $z=0$, a contradiction. Hence, $z^1>0$. Let $i\in[d]\setminus\{1\}$ and suppose that $z^i=0$. Then, the same inequality enforces $(1-\mu^{i1})\frac{z^1}{S^1_t(\o)}=0$, which yields $\mu^{i1}=1$ as $z^1>0$. We get a contradiction to our assumption that $\mu^{i1}\in [0,1)$, whence $z^i>0$. Hence, $z\in \hR^d_{++}$.
\qed 

Next we introduce the notion of {\it consistent price processes } in an analogous way to its discrete-time counterpart given by Schachermayer \cite{Sch}. 

\begin{defn}
\label{Consistent}
A process $(Z_t)_{t\in[0,T]}\in\hL^1_\hF([0,T]\times\O,\hR^d)$ is called a \textbf{consistent price process} for the solvency cone process $(\hat K_t)_{t\in[0,T]}$ if $Z$ is a $\hP$-martingale under and $\hP\{Z_t\in \hat{K}^+_t \setminus \{0\}\}=1$ for each $t\in[0,T]$. 
\end{defn}

In what follows, for a consistent price process $(Z_t)_{t\in[0,T]}$, we define $R^j_t:=\frac{\hat{Z}^j_t}{\hat{Z}^1_t}$ for each $t\in [0,T]$ and $i\in[d]$.

\begin{thm}
\label{Cprice}
Let $(Z_t)_{t\in[0,T]}\in \hL^1_\hF([0,T]\times\O,\hR^d)$ be a consistent price process with $Z^1_0=1$. Let  $(V_t)_{t\in[0,T]}$ be a self-financing portfolio process as in Proposition~\ref{Vhat}. Then, the process $M_t:=\frac{1}{S^1_t}(V^1_t+\sum_{j=2}^dR^j_tV^j_t)$, $t\in[0,T]$, is a $\hP^1$-supermartingale, where   $\frac{d\hP^1}{d\hP}\big|_{\cF_T}:= Z^1_T $. 
\end{thm} 

{\it Proof}. Since we assume that $\hF$ is the standard Brownian filtration corresponding to $W$, the process $Z$ has a continuous modification, which we also denote by $Z$ with slight abuse of notation. Then, by Corollary~\ref{cor:positive}, we have $Z_t^i>0$ for every $t\in[0,T]$ and $i\in[d]$, $\hP$-a.s. Let us fix $i\in[d]$ and consider the local martingale $Y^i_t:=\int_0^t (Z^i_s)^{-1} dZ^i_s$, 
$t\in[0,T]$. Then, by the local martingale representation theorem (cf., e.g., \cite[Theorem~4.2]{karatzas}), there exists  $\eta^i\in\hL^0_\hF([0,T]\times \O,\hR^m)$ with $\hP\{\int_0^T|\eta^i_t|^2dt <\infty\}=1$, such that $Y^i_t=\int_0^t(\eta^i_s)^{\mathsf{T}}dW_s$, $t\in[0,T]$, $\hP$-a.s. In other words, $Z^i$ satisfies the linear stochastic differential equation 
\bea
\label{SDEZ}
dZ^i_t=Z^i_t(\eta^i_t)^{\mathsf{T}}dW_t, \qq t\in[0,T].
\eea
Hence, it can be written as the Dol\'eans-Dade stochastic exponential
\[
Z^i_t=Z^i_0\exp\of{\int_0^t(\eta^i_s)^{\mathsf{T}}dW_s-\frac{1}{2}\int_0^t |\eta^i_s|^2ds}, \qq t\in[0,T].
\]

Next,  note that $Z^1_0=1$, $\frac{d\hP^1}{d\hP}\big|_{\cF_T}=Z^1_T$ defines a new probability measure $\hP^1$. Then, by Girsanov Theorem, under $\hP^1$, $W^1_t=
W_t-\int_0^t\eta^1_sds$, $t\in[0,T]$, is a Brownian motion. Furthermore, it is easy to check that for any $\hF$-adapted process $M=(M_t)_{t\in[0,T]}$, $M$ is
a $\hP^1$-supermartingale if and only if $Z^1M=(Z^1_tM_t)_{t\in[0,T]}$ is a $\hP$-supermartingale.

Let $(V_t)_{t\in[0,T]}$ be a self-financing portfolio process as in Theorem~\ref{Vhat} and consider its physical units form $\hat{V}$ defined by \eqref{eq:hat}. Since, by definition, $R^j_t=\frac{\hat Z^j_t}{\hat Z^1_t}=\frac{Z^j_t}{Z^1_t}\frac{S^1_t}{S^j_t}$, we see that
\[
M_t:=\frac1{S^1_t}\Big(V^1_t+\sum_{j=2}^d R^j_tV^j_t\Big)=\hat V^1_t+\frac1{S^1_t}\sum_{j=2}^d\frac{Z^j_t}{Z^1_t}\frac{S^1_t}{S^j_t}V^j_t=\frac1{Z^1_t}Z_t^{\mathsf{T}} \hat V_t, \q t\in[0,T].
\]
Thus, to show that $M$ is a $\hP^1$-supermartingale, it suffices to show that  $Z^1M=Z^{\mathsf{T}}\hat{V}$ is a $\hP$-supermartingale. To this end, we recall the dynamics of $\hat V$ in \eqref{dhatV} and the SDEs for $Z$ in \eqref{SDEZ}. Applying It\^o's formula, we have
\[
\left\{\ba{lll}
\dis d(Z^1_t\hat{V}^1_t)=-Z^1_t\frac1{S^1_t}\langle \Th_t,\Pi\rangle dt +\hat{V}^1_t Z^1_t(\eta^1_t)^{\mathsf{T}}dW_t,\ms \\
\dis d(Z^j_t\hat{V}^j_t)=-Z^j_t\frac1{S^j_t}\langle \Th_t-\Th_t^{\mathsf{T}}, E_i\rangle dt  +\hat{V}^j_t  Z^j_t(\eta^j_t)^{\mathsf{T}} dW_t, \q j\in[d]\setminus\{1\},
\ea\right.
\]
for some $\Th\in \hS^2_{\hF}(\hM^d_+)$. Then, we deduce that
\beaa
d(Z_t^{\mathsf{T}}\hat{V}_t)=\sum_{j=1}^d d(Z^j_t\hat V^j_t)=-Z_t^{\mathsf{T}} \hat k_t dt +\sum_{j=1}^d \hat V^j_tZ^j_t(\eta^j_t)^{\mathsf{T}} dW_t, \qq t\in[0,T],
\eeaa
where $\hat k_t=(\frac{k^1_t}{S^1_t},\ldots, \frac{k^d_t}{S^d_t})^{\mathsf{T}}$ with  $k^1_t=\langle \Th_t,\Pi\rangle $ and $k^i_t=\langle \Th_t-\Th_t^{\mathsf{T}},E_i\rangle $, $ i\in[d]\setminus\{1\}$. Since $k_t=(k^1_t, \ldots, k^d_t)\in K(\Pi)$ by definition, $\hat k_t\in \hat K_t=\hat K_t(\Pi)$. Therefore, by Definition~\ref{Consistent} and $Z$ being a consistent process process, we have $Z_t\in K^+_t\setminus\{0\}$, that is, $Z_t^{\mathsf{T}} \hat k_t\ge 0$, $t\in[0,T]$, $\hP$-a.s. In other words, $Z^{\mathsf{T}}\hat V$ is a $\hP$-supermartingale, proving the theorem. 
\qed

\section{Functional Formulation of the Superhedging Problem}\label{sec:superhedging}
\setcounter{equation}{0}

In this section, using the set-valued solvency cone process $(\hat{K}_t)_{t\in[0,T]}$ defined in \eqref{eq:Khat}, we shall describe a basic formulation of the continuous-time superhedging problem for the financial model described in Section~\ref{sec:market}.

\ms 
\no {\bf Superhedging Sets and  Set-Valued Risk Measures.} The formulation is based on the direct meaning of ``superhedging" a multi-asset payoff, i.e., finding a self-financing portfolio process whose terminal value exceeds the given payoff in every component. More precisely, in view of Theorem~\ref{Vhat}, we say that a portfolio $\xi\in\hL^2_{\cF_t}(\hR^d)$ superhedges a risky position $X\in\hL^2_{\cF_T}(\hR^d)$ at time $t\in [0,T]$ if there exists $\hat{k}\in\hat{\hS}^2_{\hF}(\hat{K})$ such that $\xi-\int_t^T \hat{k}_rdr\geq X$ $\hP$-a.s. Hence, the set of all superhedging portfolios at time $t$ is given by
\bea\label{eq:SHPf}
SHP_t(X):=\cb{\xi\in\hL^2_{\cF_t}(\hR^d)\colon \exists \hat{k}\in\hat{\hS}^2_{\hF}(\hat{K}),\ \xi-\int_t^T \hat{k}_r dr\geq X}.
\eea
With the notation of Section~\ref{sec:prelim}, this definition can be rewritten as
\[
SHP_t(X)=\cb{\xi\in\hL^2_{\cF_t}(\hR^d)\colon X\in \xi+J_{t,T}[\hat{\hS}^2_{\hF}(-\hat{K})]-\hL^2_{\cF_T}(\hR^d_+)},
\]
which is the continuous-time analog of the set in \eqref{SHP}. Similar to the discrete-time case, we also define the corresponding superhedging risk measure at time $t$ by 
\bea\label{RMdefn}
R_t(X):=SHP_t(-X)=\cb{\xi\in\hL^2_{\cF_t}(\hR^d)\colon X+\xi \in J_{t,T}[\hat{\hS}^2_{\hF}(\hat{K})]+\hL^2_{\cF_T}(\hR^d_+)},
\eea 
and, for $0\leq t\leq u\leq T$, its (stepped) acceptance sets are given by
\[
A_t:=\{X\in\hL^2_{\cF_T}(\hR^d)\colon 0\in R_t(X)\}=J_{t,T}[\hat{\hS}^2_{\hF}(\hat{K})]+\hL^2_{\cF_T}(\hR^d_+),\quad A_{t,u}:=A_t\cap \hL^2_{\cF_u}(\hR^d).
\]

To avoid some degenerate cases of the superhedging sets, we work under the following assumption throughout this section:

\begin{assum}\label{asmp}
	There exists a consistent price process $(Z_t)_{t\in[0,T]}\in \hL^1_{\hF}([0,T]\times\O,\hR^d)$ with $Z_0^1=1$.
	\end{assum}

\begin{prop}\label{RMdirect}
	The family $(R_t)_{t\in[0,T]}$ of set-valued functions defined by \eqref{RMdefn} is a conditionally coherent and normalized dynamic set-valued risk measure.
\end{prop}

{\it Proof.} Let $t\in [0,T]$. We first show that the function $R_t$ takes values in $\sP_+(\hL^2_{\cF_t}(\hR^d))$. To that end, let us fix $X\in \hL^2_{\cF_T}(\hR^d)$ and show that $R_t(X)= R_t(X)+ \hL^2_{\cF_t}(\hR^d_+)$. The inclusion $\subseteq$ is trivial since $0\in\hL^2_{\cF_t}(\hR^d_+)$. Conversely, let $\xi\in R_t(X)$ and $\eta\in  \hL^2_{\cF_t}(\hR^d_+)$. Then, $X+\xi\in J_{t,T}[\hat{\hS}^2_{\hF}(\hat{K})]+\hL^2_{\cF_T}(\hR^d_+)$, and therefore
\beaa 
X+\xi+\eta &\in& J_{t,T}[\hat{\hS}^2_{\hF}(\hat{K})]+\hL^2_{\cF_T}(\hR^d_+)+\hL^2_{\cF_t}(\hR^d_+)\\
&\subseteq & J_{t,T}[\hat{\hS}^2_{\hF}(\hat{K})]+\hL^2_{\cF_T}(\hR^d_+)+\hL^2_{\cF_T}(\hR^d_+)\\
&=&J_{t,T}[\hat{\hS}^2_{\hF}(\hat{K})]+\hL^2_{\cF_T}(\hR^d_+)
\eeaa 
since $\hL^2_{\cF_T}(\hR^d_+)$ is a convex cone. Hence, $\xi+\eta\in R_t(X)$ so that $R_t(X)+\hL^2_{\cF_T}(\hR^d_+)\subseteq R_t(X)$.

Next, we check the properties of conditional set-valued risk measures. First, let $X,Y\in \hL^2_{\cF_T}(\hR^d)$ with $Y\geq X$, i.e., $Z:=Y-X\in \hL^2_{\cF_T}(\hR^d_+)$. Let  $\xi\in R_t(X)$. Then, $\xi\in\hL^2_{\cF_t}(\hR^d)$ and $X+\xi=Y-Z+\xi\in J_{t,T}[\hat{\hS}^2_{\hF}(\hat{K})]+\hL^2_{\cF_T}(\hR^d_+)$. Then,
\[
Y+\xi\in Z+J_{t,T}[\hat{\hS}^2_{\hF}(\hat{K})]+\hL^2_{\cF_T}(\hR^d_+)\subseteq J_{t,T}[\hat{\hS}^2_{\hF}(\hat{K})]+\hL^2_{\cF_T}(\hR^d_+).
\]
Hence, $\xi\in R_t(Y)$ so that $R_t(X)\subseteq R_t(Y)$, i.e., $R_t$ is monotone.

Let $X\in \hL^2_{\cF_T}(\hR^d)$ and $\eta\in \hL^2_{\cF_t}(\hR^d)$. Then, we have
\beaa 
R_t(X+\eta)&=&\cb{\xi\in\hL^2_{\cF_t}(\hR^d)\colon X+\eta +\xi \in J_{t,T}[\hat{\hS}^2_{\hF}(\hat{K})]+\hL^2_{\cF_T}(\hR^d_+)}\\
&=& \cb{\xi^{\prime}\in\hL^2_{\cF_t}(\hR^d)\colon X+\xi^{\prime} \in J_{t,T}[\hat{\hS}^2_{\hF}(\hat{K})]+\hL^2_{\cF_T}(\hR^d_+)}-\eta\\
&=& R_t(X)-\eta.
\eeaa 
Hence, $R_t$ is translative.

Note that $R_t(0)=A_t\cap \hL^2_{\cF_t}(\hR^d)$. Since $0\in A_t$, we have $R_t(0)\neq\emptyset$.  Next, we show that $R_t(0)\neq \hL^2_{\cF_t}(\hR^d)$. To get a contradiction, suppose that $
\hL^2_{\cF_t}(\hR^d)\subseteq A_t = J_{t,T}[\hat{\hS}^2_{\hF}(\hat{K})]+\hL^2_{\cF_T}(\hR^d_+)$. Let us fix $\xi\in \hL^2_{\cF_t}(\hR^d)$. Then, there exists $\hat{k}^{\xi}\in\hat{\hS}^2_{\hF}(\hat{K})$ such that $\xi\geq \int_t^T \hat{k}^{\xi}_rdr$. For each $u\in[t,T]$, let $\hat{V}^{\xi}_u:=\xi-\int_t^u \hat{k}^{\xi}_rdr$. Then, $(\hat{V}_u)_{u\in[t,T]}$ is a self-financing portfolio process expressed in physical units over the interval $[t,T]$; see Theorem~\ref{Vhat}(v). Let $Z$ be a consistent price process wit $Z^1_0=1$, whose existence is guaranteed by Assumption~\ref{asmp}. Then, applying Theorem~\ref{Cprice} over the interval $[t,T]$, we obtain that $Z^{\mathsf{T}}\hat{V}$ is a $\hP$-supermartingale. In particular,
\[
Z^{\mathsf{T}}_t \xi= Z^{\mathsf{T}}_t\hat{V}_t\geq  \hE[Z^{\mathsf{T}}_T\hat{V}_T | \cF_t]=\hE\sqb{Z_T^{\mathsf{T}}\of{\xi-\int_t^T \hat{k}^{\xi}_rdr}}\geq 0,
\]
where the last inequality follows since $\xi\geq \int_t^T \hat{k}^{\xi}_rdr$ and $Z_T\geq 0$ by Corollary~\ref{cor:positive}. The same corollary also yields that $Z_t\in \hR^d_{++}$ $\hP$-a.s. Since $\xi\in\hL^2_{\cF_t}(\hR^d)$ is arbitrary, from the above calculations, we get $-\infty = \essinf_{\xi\in \hL^2_{\cF_t}(\hR^d)}Z^{\mathsf{T}}_t\xi \geq 0$, which is a contradiction. Hence, $R_t(0)\neq \hL^2_{\cF_t}(\hR^d)$ so that $R_t$ is finite at zero.

Let $X,Y\in\hL^2_{\cF_T}(\hR^d)$ and $\alpha,\beta\in\hL^\infty_{\cF_t}(\hR_+)$. We claim that
\bea\label{eq:coneconv}
\alpha R_t(X)+\beta R_t(Y)\subseteq R_t(\alpha X+\beta Y).
\eea
Let $\xi\in R_t(X)$ and $\eta\in R_t(Y)$. Hence, $X+\xi,Y+\eta\in J_{t,T}[\hat{\hS}^2_{\hF}(\hat{K})]+\hL^2_{\cF_T}(\hR^d_+)$. Since $\alpha ,\beta$ are nonnegative $\cF_t$-measurable random variables and $\hat{K}$ is a convex cone-valued process, we have $\alpha J_{t,T}[\hat{\hS}^2_{\hF}(\hat{K})]+\beta J_{t,T}[\hat{\hS}^2_{\hF}(\hat{K})]=J_{t,T}[\hat{\hS}^2_{\hF}(\hat{K})]$. It follows that $\alpha(X+\xi)+\beta (Y+\eta)\in J_{t,T}[\hat{\hS}^2_{\hF}(\hat{K})]+\hL^2_{\cF_T}(\hR^d_+)$. Hence, $\alpha  \xi + \beta \eta \in R_t(\alpha X+\beta Y)$. Taking $\alpha=\lambda$ and $\beta=1-\lambda$ for each $\lambda\in \hL^\infty_{\cF_t}([0,1])$, we see that $R_t$ is conditionally convex. Taking $\beta=0$, we obtain $\alpha R_t(X)\subseteq R_t(\alpha X)$ for every $\alpha\in\hL^{\infty}_{\cF_t}(\hR_+)$. Then, assuming that $\alpha\in \hL^\infty_{\cF_t}(\hR_{++})$, we have
\[
R_t(\alpha X)=\alpha\frac{1}{\alpha}R_t(\alpha X)\subseteq \alpha R_t(X)
\]
as well. Hence, $R_t(\alpha X)=\alpha R_t(X)$ so that $R_t$ is conditionally positively homogeneous.

Let $X\in L^2_{\cF_T}(\hR^d)$. Since $0\in R_t(0)$, we have  $R_t(X)\subseteq R_t(X)+R_t(0)$. By \eqref{eq:coneconv}, we also have $R_t(X)+R_t(0)\subseteq R_t(X)$. Hence, $R_t$ is normalized.

Thus, $(R_t)_{t\in[0,T]}$ is a conditionally coherent and normalized dynamic set-valued risk measure.
\qed

\ms 
\no {  \bf The Functional Dynamic Programming Principle.} The next proposition states the time-consistency of the family $(R_t)_{t\in[0,T]}$ in the set-valued setting.

\begin{prop}\label{RMdirect-MPTC}
	The set-valued dynamic risk measure $(R_t)_{t\in[0,T]}$ is multi-portfolio time-consistent.
\end{prop}

{\it Proof.} Since $(R_t)_{t\in [0,T]}$ is normalized, by Theorem~\ref{ConRm}, multi-portfolio time-consistency is equivalent to $A_t=A_{t,u}+A_u$ for every $0\leq t<u\leq T$, which we check next. Note that
\[
A_t=J_{t,T}[\hat{\hS}^{2}_{\hF}(\hat{K})]+\hL^2_{\cF_T}(\hR^d_+),\quad A_{t,u}=A_t\cap \hL^2_{\cF_u}(\hR^d).
\]
Let $\xi\in A_{t,u}$ and $X\in A_{u}$. Then $\xi\in \hL^2_{\cF_u}(\hR^d)$ and $\xi\geq \int_t^T \hat{k}^{\xi}_rdr$ for some $ \hat{k}^{\xi}\in \hat{\hS}^2_\hF(\hat{K})$, and similarly, $X\in \hL^2_{\cF_T}(\hR^d)$ and $\eta\geq \int_u^T \hat{k}^{X}_rdr$ for some $ \hat{k}^{X}\in \hat{\hS}^2_\hF(\hat{K})$. Thus, $Y:=\xi+X\in \hL^2_{\cF_T}(\hR^d)$ and $Y\geq \int_t^T \hat{k}^{\xi}_rdr+\int_u^T \hat{k}^{X}_rdr =\int_t^T (\hat{k}^{\xi}_r+\1_{[u,T]}(r)\hat{k}^{X}_r)dr$, and therefore $Y\in A_t$. 

Conversely, let $X\in A_t$. Then $X\in \hL^2_{\cF_T}(\hR^d)$  and $X\geq \int_t^T \hat{k}_rdr$ for some $\hat{k}\in \hat{\hS}^2_\hF(\hat{K})$.  Let us take $\xi:=\int_t^u\hat{k}_rdr$ and $Y:=X-\xi$. Then, clearly $\xi\in A_{t,u}$. On the other hand, $Y\geq \int_u^T \hat{k}_rdr$, and hence $Y\in A_u$. Since $X=\xi+Y$, the result follows. 
\qed 

	\begin{rem}\label{rem:Gconsis}
	{\rm As an immediate consequence of Proposition~\ref{RMdirect-MPTC}, the family $(SHP_t)_{t\in [0,T]}$ satisfies the following recursive property:
		\beaa
		SHP_t(X)=R_t(-X)= \bigcup _{\eta\in R_u(-X)}R_t(-\eta)=\bigcup _{\eta\in SHP_u(X)}SHP_t(\eta), 
		\eeaa
		for every $0\leq t\leq u\leq T$ and $X\in \hL^2_{\cF_T}(\hR^d)$.\qed 
		}
	\end{rem}
	
	We now sharpen the observation in Remark~\ref{rem:Gconsis} and obtain a more useful recursive relation for the family $(SHP_t)_{t\in[0,T]}$, which can be seen as a set-valued dynamic programming principle.
	
	\begin{thm}\label{G1recur}
	Let $0\leq t\leq u\leq T$ and $X\in \hL^2_{\cF_T}(\hR^d)$. Then, it holds
	\bea
	\label{SHPt}
	SHP_t(X)=\big(SHP_u(X)+J_{t,u}[ \hat{\hS}^2_\hF(\hat{K})]\big)\cap \hL^2_{\cF_t}(\hR^d).
	\eea
	\end{thm}
	
	{\it Proof}. For a random vector $\eta\in \hL^2_{\cF_u}(\hR^d)$, first note that $\eta\in SHP_u(X)$ if and only if there exists $\tilde{k}\in \hat{\hS}^2_\hF(\hat{K})$ such that $\eta\geq X+J_{u,T}(\tilde{k})$. Thus, by Remark~\ref{rem:Gconsis}, we have
	\begin{equation}\label{Gconsis1}
		\begin{aligned}
			SHP_t(X)
			&=\big\{\xi\in \hL^2_{\cF_t}(\hR^d)\colon\exists \eta\in SHP_u(X), \xi\in SHP_t(\eta) \big\} \\
			&=\big\{\xi\in \hL^2_{\cF_t}(\hR^d)\colon\exists \hat{k}\in \hat{\hS}^2_\hF(\hat{K}), \exists \eta\in SHP_u(X), \xi\geq \eta+J_{t,T}(\hat{k}) \big\} \\
			&\subseteq \big\{\xi\in \hL^2_{\cF_t}(\hR^d)\colon\exists \hat{k},\tilde{k} \in \hat{\hS}^2_\hF(\hat{K}), \xi\geq X+J_{u, T}(\tilde{k})+J_{t,T}(\hat{k}) \big\} \\
			&=\big\{\xi\in \hL^2_{\cF_t}(\hR^d)\colon\exists \hat{k},\tilde{k} \in 
			\hat{\hS}^2_\hF(\hat{K}), \xi-J_{t,u}(\hat{k})\geq X+J_{u,T}(\tilde{k}+\hat{k}) \big\}.
		\end{aligned}
	\end{equation}
	Since $\hL^2_{\cF_t}(\hR^d)\subseteq \hL^2_{\cF_u}(\hR^d)$, having $\xi-J_{t,u}(\hat{k})\geq X+J_{u,T}(\tilde{k}+\hat{k})$ for some $\xi\in \hL^2_{\cF_t}(\hR^d)$ implies that $ \xi-J_{t,u}(\hat{k})\in SHP_u(X)$. Hence, from \eqref{Gconsis1}, we obtain
	\begin{equation*}
		\begin{aligned}
			SHP_t(X)&\subseteq \big\{\xi \in \hL^2_{\cF_t}(\hR^d)\colon \exists \hat{k}\in\hat{\hS}^2_{\hF}(\hat{K}),\ \xi -J_{t,u}(\hat{k})\in SHP_u(X)\big\}\\
			&=\big(SHP_u(X)+J_{t,u}[ \hat{\hS}^2_\hF(\hat{K})]\big)\cap \hL^2_{\cF_t}(\hR^d).
		\end{aligned}
	\end{equation*}
	Conversely, let $\xi \in (SHP_u(X)+J_{t,u}[ \hat{\hS}^2_\hF(\hat{K})])\cap \hL^2_{\cF_t}(\hR^d)$.  Then, there exist $\eta \in SHP_u(X)$ and $\hat{k}\in \hat{\hS}^2_\hF(\hat{K})$ such that $\xi=\eta+J_{t,u}(\hat{k})$. Since $\eta \in SHP_u(X)$, there exists $\tilde{k}\in \hat{\hS}^2_{\hF}(\hat{K})$ such that $\eta\geq X+J_{u,T}(\tilde{k})$. Hence, $\xi \geq X+J_{u,T}(\tilde{k})+J_{t,u}(\hat{k})=X+J_{t,T}(\1_{[t,u)}\hat{k}+\1_{[u,T]} \tilde{k}) $ so that $\xi\in SHP_t(X)$. 
	\qed

\begin{rem}
		{\rm  We note that the recursive relation in Theorem~\ref{G1recur} is closely related to the notion of \emph{conditional core} introduced in \cite{LepMol}. While their definition applies for random closed sets, we now give a slightly modified version of the concept that fits our case. 
			Let $M\subseteq \hL^2_{\cF}(\hR^d)$ be a nonempty set  
			and $\cG\subseteq\cF$ be a sub-$\si$-field of $\cF$. We define the {\it $\cG$-conditional core} $\bm[M|\cG]$ of $M$ to be the largest $\cG$-decomposable closed set $M^\prime\subseteq \hL^2_{\cG}(\hR^d)$ such that $M^\prime\subseteq \ol{\dec}_\cG (M)$. It can be argued that (cf. \cite{LepMol})  if $M$ is convex, or a cone, then so is $\bm[M|\cG]$, whenever it exists.  Now for any closed subset $M\subseteq \hL^2_{\cF_u}(\hR^d)$, 
			we consider $M_t:=\hL^2_{\cF_t}(\hR^d)\cap M$, $t \leq u$.  Then, clearly $M_t=\bm[M|\cF_t]$ provided that $M_t$ is 
			$\cF_t$-decomposable. Since $SHP_t(X)$ is $\cF_t$-decomposable, $t\in[0,T]$, we may write \eqref{SHPt} in Theorem~\ref{G1recur} as the following {\it dynamic programming principle} in terms of the conditional core:
			\bea\label{Grecur1}
			SHP_t(X)=\bm\big[SHP_u(X)+ J_{t,u}[ \hat{\hS}^2_\hF(\hat{K})]\big|\cF_t\big],
			\eea
			where  $0\le t\le u$ and $X\in\hL^2_{\cF_T}(\hR^d)$.
			\qed}
\end{rem}

\begin{rem}\label{rem:lebesgue}
	{\rm		As an alternative to the formulation in this section, we may consider the following relaxed formulation of the superhedging problem based on the set-valued Lebesgue integral reviewed in Section~\ref{sec:prelim}. Given a financial position $X\in \hL^2_{\cF_T}(\hR^d)$, we may say that a portfolio $\xi\in\hL^2_{\cF_t}(\hR^d)$ superhedges $X$ at time $t\in[0,T]$ if $\xi\in X+ \hS^2_{\cF_T}(\int_t^T \hat{K}_rdr)$, yielding the superhedging risk measure 
		\[
		R^\prime_t(X):=\cb{\xi\in\hL^2_{\cF_t}\colon \xi+X\in \hS^2_{\cF_T}\of{\int_t^T \hat{K}_rdr}}.
		\]
		Notably, $R_t^\prime$ possesses some of the nice features of $R_t$ such as translativity and convexity, and the corresponding acceptance set
		\[
		A_t^\prime:=\{X\in\hL^2_{\cF_T}(\hR^d)\colon 0\in R^\prime_t(X) \}=\hS^2_{\cF_T}\of{\int_t^T \hat{K}_rdr}=\ol{\dec}_{\cF_T}(J_{t,T}[\hS^2_{\hF}(\hat{K})])
		\]
		is closed in $\hL^2_{\cF_t}(\hR^d)$. However, this definition lacks a clear financial interpretation. In fact, an acceptable position $X\in A^\prime_t$ can only be written as the limit of a sequence of positions of the form $\sum_{i=1}^m {\bf 1}_{B_i}\int_t^T \hat{k}^i_rdr$, where $(B_i)_{i\in[m]}$ is an $\cF_T$-partition of $\O$ and $\hat{k}^1,\ldots,\hat{k}^m\in \hS^2_{\hF}(\hat{K})$. Such an approximation of $X$ does not correspond to a clear superhedging strategy since $\sum_{i=1}^m {\bf 1}_{B_i}\hat{k}^i$ is not adapted in general. Hence, among the two notions of set-valued integration with respect to the time variable, we conclude that the functional set-valued integral is a more suitable alternative for the superhedging problem.
		\qed}
	\end{rem}

\section{Pathwise Formulation of the Superhedging Problem}
\label{sec:pathwise}
\setcounter{equation}{0}

In this section, we will provide a pathwise formulation of the superhedging problem based on the canonical space of continuous functions.

\ms 
\no {\bf Motivation.} Let us briefly explain the motivation to pass to the path-space setting. According to the functional formulation of the superhedging problem in Section~\ref{sec:superhedging}, the set of superhedging strategies of a risky position $X\in \hL^2_{\cF_T}(\hR^d)$ at time $t\in[0,T]$ is defined as a convex cone $R_t(X)\subseteq \hL^2_{\cF_t}(\hR^d)$. In view of Remark~\ref{rem:drm}, the set $R_t(X)$ is also $\cF_t$-decomposable and
\[
\cl_{\hL^2}(R_t(X))=\hS^2_{\cF_t}(\tilde{R}_t(X))
\]
for some set-valued random variable $\tilde{R}_t(X)\in\sA^p_{\cF_t}(\O,\sG(\hR^d))$. Intuitively, one would like to treat the set-valued process $(t,\o)\mapsto \tilde{R}_t(X)(\o)$ as a pathwise version of the collection $(R_t(X))_{t\in[0,T]}$ and expect that this set-valued process satisfies a local form of the multi-portfolio time-consistency in Proposition~\ref{RMdirect-MPTC} and a local form of the dynamic programming principle in Theorem~\ref{G1recur}. However, an attempt to prove a dynamic programming principle directly for the set-valued process $(t,\o)\mapsto \tilde{R}_t(X)(\o)$ involves numerous technical challenges such as joint measurability and the lack of decomposability for the functional set-valued integral. By formulating superhedging sets as set-valued random variables on the path space, we will prove a pathwise version of the dynamic programming principle for a slightly relaxed version of $(t,\o)\mapsto \tilde{R}_t(X)(\o)$. 

\ms
\no {\bf Shifted and Concatenated Paths.} Throughout this section, we will work under the following filtered probability space. Let us take $\O=\hC_0([0,T],\hR^m)$ be the collection of all continuous functions $\o=(\o_t)_{t\in[0,T]} \colon  [0,T]\to \hR^m$ with $\o_0=0$, equipped with the norm $\o\mapsto \|\o\|_\infty:=\sup_{t\in[0,T]}|\o_t|$ for uniform convergence. We take $\cF$ to be the Borel $\sigma$-algebra on $\O$ and $\hP$ to be the Wiener measure on $(\O,\cF)$. Let $W=(W_t)_{t\in[0,T]}$ denote the canonical process, i.e., $W_t(\o):=\o_t$ for every $t\in[0,T]$ and $\o\in\O$. We take $\hF$ to be the standard filtration of $W$ that satisfies the usual conditions.

Following \cite{dynamicgames,meanfield}, we will work with concatenations of paths in $\O$. Let $t\in [0,T]$. For each $\o,\tilde{\o}\in\O$, we define their concatenation $\o\oplus_{t} \tilde{\o}\in\O$ at $t$ by
\[
(\o\oplus_{t} \tilde{\o})_r:= \o_r {\bf{1}}_{[0,t)}(r)+(\o_{t}+\tilde{\o}_{r-t}){\bf{1}}_{[t,T]}(r),\quad r\in [0,T].
\]
Since $(\O,\cF)$ is a standard measurable space, there exists a regular conditional probability $(\hP^{\o}_t)_{\o\in\O}$ given $\cF_{t}$. Without loss of generality, we assume that $\hP^{\o}_{t}(\{\tilde{\o}\in\O\colon \o_s=\tilde{\o}_s\text{ for every }s\in[0,t]\})=1$ for every $\o\in\O$. For each $\o\in\O$, $\xi\in \hL^0_{\cF}(\hR^d)$, we define
\[
\xi^{t,\o}(\tilde{\o}):=\xi(\o\oplus_{t}\tilde{\o}),\quad \tilde{\o}\in\O;\quad\quad \hP^{t,\o}(A):=\hP^{\o}_{t}(\o\oplus_{t}A),\quad A\in\cF,
\] 
where $\o\oplus_{t}A:=\{\o\oplus_{t}\tilde{\o}\colon \tilde{\o}\in A\}$. Note that $\hP^{t,\o}$ is a probability measure on $(\O,\cF)$ for each $\o\in\O$. Indeed, by the Markov property of Wiener process, we have $\hP^{t,\o}=\hP$ for $\hP$-a.e. $\o\in\O$. Moreover, whenever $\xi\in \hL^1_{\cF}(\hR^d)$, it is easy to verify that
\bea\label{eq:tomega}
\hE[\xi|\cF_{t}](\o)=\hE^{\o}_{t}[\xi]=\hE^{t,\o}[\xi^{t,\o}]=\hE[\xi^{t,\o}]
\eea
for $\hP$-a.e.~$\o\in\O$. We also define the time-shifted Wiener process $W^t=(W^t_u)_{u\in[t,T]}$ by $W^{t}_u:=W_{u-t}$ for each $u\in [t,T]$, and denote by $\hF^t=(\cF^{t}_u)_{u\in[t,T]}$ its natural filtration, i.e., $\cF^t_u=\sigma(\{W^t_r\colon\ r\in [t,u]\})$ for each $u\in [t,T]$. Note that, for each $\o,\tilde{\o}\in\O$ and $u\in[t,T]$, 
\[
W^{t,\o}_u(\tilde{\o}):=(W_u)^{t,\o}(\tilde{\o})=W_u(\o\oplus_t\tilde{\o})=\o_t+\tilde{\o}_{u-t}=W_t^{t,\o}(\tilde{\o})+W^t_u(\tilde{\o}),
\]
i.e., $W^t_u=W^{t,\o}_u-W^{t,\o}_t$. Since the increments of $(W^{t,\o}_u)_{u\in[t,T]}$ do not depend on $\o$, we simply use $W^t$ in the sequel.

\ms 
\no {\bf  Shifted Dynamics.} In this setting, we will consider the stock price process $(S_t)_{t\in[0,T]}$ that is uniquely defined by the dynamics in \eqref{BS} with the additional assumption that $b\colon [0,T]\to \hR^d$ and $\sigma\colon[0,T]\to \hR^{d\times m}$ are deterministic functions of time. Given $\eta\in \hL^2_{\cF_t}(\hR^d_{++})$, we denote by $S^{\eta;t}=(S^{\eta;t}_u)_{u\in[t,T]}$ the unique solution of the same SDE starting at $t$ with initial value $\eta$, i.e.,
\[
dS^{\eta;t}_u=\diag(S^{\eta;t}_u)(b_udu+\sigma_udW_u),\ u\in[t,T];\quad S^{\eta;t}_t=\eta.
\]
In particular, by the flow property of the SDE, we have $S^{S_t;t}_u=S_u$ for every $u\in[t,T]$ $\hP$-a.s.

Let $\o\in\O$. Then, for $\hP$-a.e. $\tilde{\o}\in\O$, we have
\begin{equation*}
	\begin{aligned}
		S^{t,\o}_u(\tilde{\o})&=S_u(\o\oplus_{t}\tilde{\o})\\
		&=S_t(\o\oplus_t\tilde{\o})+\of{\int_{t}^u\diag(S_r)b_rdr}(\o\oplus_t\tilde{\o})+\of{\int_{t}^u \diag(S_r)\sigma_rdW_r}(\o\oplus_t\tilde{\o})\\
		&=S_t(\o)+\of{\int_{t}^u\diag(S^{t,\o}_r)b_rdr}(\tilde{\o})+\of{\int_{t}^u \diag(S^{t,\o}_r)\sigma_rdW^{t}_r}(\tilde{\o})
		\end{aligned}
		\end{equation*}
for every $u\in [t,T]$. Hence, the shifted stock price dynamics can be written as
\bea\label{eq:shifted}
dS_u^{t,\o}=\diag(S_u^{t,\o})(b_udu+\sigma_udW^t_u),\ u\in[t,T],\quad S^{t,\o}_t=S_t(\o).
\eea

\ms 
\no {\bf Approximate Solvency Cones.} Recall the deterministic solvency cone $K(\Pi)$ defined in \eqref{KPi}. For some formulations in this section, we will rely on the following \emph{halfspace representation} of $K(\Pi)$. Since $K(\Pi)$ is a polyhedral closed convex cone satisfying $K(\Pi)=K(\Pi)+\hR^d_+$ (see Proposition~\ref{prop:cone-upper}), we may find direction vectors $w^1,\ldots,w^N\in \hR^d_+\setminus\{0\}$ for some $N\in\hN$ such that
\[
K(\Pi)=\bigcap_{n=1}^N \{x\in\hR^d\colon (w^n)^{\mathsf{T}}x\geq 0\}.
\]
Without loss of generality, we assume that $|w^n|=1$ for each $n\in[N]$. Given $\varepsilon\geq 0$, we introduce the $\varepsilon$-solvency cone in physical units at $y\in\hR^d_{++}$ by
\[
\hat{\sK}^{\varepsilon}(y):=\bigcap_{n=1}^N \{x\in\hR^d\colon (\diag(y)w^n)^{\mathsf{T}}x\geq -\varepsilon |x|_\infty\},
\]
which is a polyhedral closed convex cone; here, $|x|_\infty:=\max_{i\in[d]}|x_i|$ for $x\in\hR^d$. Note that
\[
\hat{\sK}(y):=\hat{\sK}^0(y)=\bigcap_{n=1}^N\{x\in\hR^d\colon (\diag(y)w^n)^{\mathsf{T}}x\geq 0\}=\diag(y)^{-1}K(\Pi).
\]
Given $(t,\o)\in[0,T]\times\O$, we also define $\hat{K}^{\varepsilon}_t(\o):=\hat{\sK}^{\varepsilon}(S_t(\o))$ and its shifted version
\[
\hat{K}_u^{\varepsilon;t,\o}(\tilde{\o}):=\hat{K}^{\varepsilon}_u(\o\oplus_t\tilde{\o})=\hat{\sK}^{\varepsilon}(S^{t,\o}_u(\tilde{\o}))
\]
for every $(u,\tilde{\o})\in[t,T]\times\O$. With this notation, the cone-valued process in \eqref{eq:Khat} can be expressed by $\hat{K}_t(\o)=\hat{K}^{0}_t(\o)$ and we write $\hat{K}^{t,\o}_u(\tilde{\o}):=\hat{K}^{0;t,\o}_u(\tilde{\o})$. 

As the process $S$ has $\hP$-a.s. continuous paths, the cone-valued process $\hat{K}^{\varepsilon}$ is uniquely defined up to indistinguishability and we can introduce process selectors of it with additional path regularity properties. To that end, let $\hD^0_{\hF}(\hR^d)\subset \hL^0_{\hF}(\hR^d)$ denote the space of all $\hF$-progressively measurable $\hR^d$-valued c\`{a}dl\`{a}g processes and define the subspace
\[
\hat{\hD}^2_{\hF}(\hR^d):=\cb{\hat{k}\in \hD^0_{\hF}(\hR^d)\colon \diag(S)\hat{k}\in \hL^2_{\hF}([0,T]\times\O,\hR^d)}.
\]
Then, the corresponding space of square-integrable portfolio processes in physical units is given by
\[
\hat{\hD}_{\hF}^{2}(\hat{K}^{\varepsilon}):=\hat{\hD}^2_{\hF}(\hR^d)\cap \hS^0_{\hF}(\hat{K}^{\varepsilon}).
\]
Note that each $\hat{k}\in \hat{\hD}^2_{\hF}(\hat{K}^{\varepsilon})$ satisfies $\hP\{\hat{k}_r\in \hat{K}_r^{\varepsilon}\ \forall r\in[0,T]\}=1$. The shifted space $\hat{\hD}^2_{\hF^t}(\hat{K}^{\varepsilon;t,\o})$ is defined similarly for each $(t,\o)\in[0,T]\times\O$.

The next lemma is a useful observation about the set-valued functions $(\hat{\sK}^\varepsilon)_{\varepsilon\geq 0}$.

\begin{lem}\label{lem:Keps}
Let $\varepsilon_1,\varepsilon_2\geq 0$ and $y,y^\prime\in\hR^d_{++}$. If $|y-y^\prime|\leq \varepsilon_1$, then $\hat{\sK}^{\varepsilon_2}(y)\subseteq \hat{\sK}^{\varepsilon_1+\varepsilon_2}(y^\prime)$.
\end{lem}

{\it Proof.} 	Suppose that $|y-y^\prime|\leq \varepsilon_1$. Let $x\in\hat{\sK}^{\varepsilon_2}(y)$ and fix some $n\in\{1,\ldots,N\}$. Note that $(\diag(y)w^n)^{\mathsf{T}}x\geq -\varepsilon_2|x|_\infty$. Then,
\beaa
(\diag(y^\prime)w^n)^{\mathsf{T}}x &=& (\diag(y^\prime-y)w^n)^{\mathsf{T}}x+(\diag(y)w^n)^{\mathsf{T}}x\\
&\geq& \inf_{z\in \hB_{\hR^d}(\varepsilon_1)} (\diag(z)w^n)^{\mathsf{T}}x-\varepsilon_2|x|_\infty\\
&=&\inf_{z\in \hB_{\hR^d}(\varepsilon_1)} (\diag(x)w^n)^{\mathsf{T}}z -\varepsilon_2|x|_\infty\\
&=& -\varepsilon_1|\diag(x)w^n| -\varepsilon_2|x|_\infty \\
&\geq& -\varepsilon_1 |x|_\infty |w^n|  -\varepsilon_2|x|_\infty= -(\varepsilon_1+\varepsilon_2) |x|_\infty,
\eeaa 
where the evaluation of the infimum follows by a simple geometric observation (or by the definition of dual norm). Hence, $x\in \hat{\sK}^{\varepsilon_1+\varepsilon_2}(y^\prime)$.
\qed 

\ms 
\no {\bf Approximate Superhedging Sets.} As a financial position to superhedge, we consider a random vector $X$ that is a deterministic function of the stock price history, i.e., we take $X:=g\circ S$ for some contract function $g\colon \hC_{s_0}([0,T],\hR^d)\to\hR^d$, where $\hC_{s_0}([0,T],\hR^d)$ is the space of continuous functions $f=(f_t)_{t\in[0,T]}\colon[0,T]\to\hR^d$ with $f_0=s_0\in \hR^d_{++}$, equipped with the supremum norm $\|\cdot\|_\infty$. We assume that $g$ satisfies the following two conditions:
\begin{enumerate}
	\item $\hE[|g\circ S|^2]<+\infty$, i.e., $X\in \hL^2_{\cF_T}(\hR^d)$.
	\item $g$ is Lipschitz continuous, i.e., there exists $L\geq 1$ such that $|g(f)-g(f^\prime)|\leq L\|f-f^{\prime}\|_\infty$ for every $f,f^\prime \in \hC_{s_0}([0,T],\hR^d)$.
	\end{enumerate}
Such $X$ covers most of the frequently used payoff structures of vanilla and exotic options.
	
Let $\varepsilon\in [0,1]$. We define the (functional) $\varepsilon$-superhedging set at time $t\in[0,T]$ as
\[
SHP_t^{\varepsilon}(X)
:=\cb{\eta \in \hL^{2}_{\cF_t}(\hR^d)\colon \exists \hat{k}\in \hat{\hD}^{2}_{\hF}(\hR^d),\
	\hP\cb{\begin{array}{@{}c@{}} \hat{k}_r\in \hat{K}_r^{\varepsilon}\ \forall r\in [t,T],\\ \eta-\int_t^T \hat{k}_rdr+ L\varepsilon{\bf1} \geq  X\end{array}
		\;\Big\vert\; \cF_t }\neg\geq\neg  1- \varepsilon}.
\]
Clearly, $SHP_t^0(X)=SHP_t(X)$ recovers the superhedging set given in \eqref{eq:SHPf}.

Finally, we define the value of the local superhedging problem as
\bea\label{eq:V}
\hV_t(\o):=\bigcap_{\varepsilon\in (0,1]}\hV^\varepsilon_t(\o),\quad (t,\o)\in [0,T]\times\O,
\eea
where $\hV^{\varepsilon}_t(\o):=\tilde{\hV}^{\varepsilon}_t(\o)+\hB_{\hR^d}(\varepsilon)$ and
\beaa
\tilde{\hV}^{\varepsilon}_t(\o):=\cb{y\in\hR^d\colon \exists \hat{k}^{\o}\in \hat{\hD}_{\hF^t}^{2}(\hR^d),\ \hP\cb{\begin{array}{@{}c@{}}\hat{k}_r^{\o}\in\hat{K}_r^{\varepsilon;t,\o}\ \forall r\in[t,T],\\ y-\int_t^T \hat{k}^{\o}_rdr+ L\varepsilon{\bf 1} \geq X^{t,\o}\end{array}}\geq 1-\varepsilon}.
\eeaa 
Note that $\hV_t^0(\o)\subseteq \hV_t(\o)$ since $\hV_t^0(\o)=\tilde{\hV}^0_t(\o)\subseteq \tilde{\hV}^\varepsilon_t(\o)\subseteq \hV^\varepsilon_t(\o)$ for every $\varepsilon\in (0,1]$. The next lemma makes this observation slightly sharper.

\begin{lem}\label{lem:V}
	Let $0\leq \varepsilon^\prime<\varepsilon\leq 1$ and $(t,
	\o)\in[0,T]\times\O$. Then, $\cl(\hV_t^{\varepsilon^\prime}(\o))\subseteq \hV_t^{\varepsilon}(\o)$.
	\end{lem}
	
{\it Proof.} Let $y\in \cl(\hV^{\varepsilon^\prime}_t(\o))$. Then, there exists $\hat{y}\in \hV_t^{\varepsilon^\prime}(\o)$ such that $|y-\hat{y}|\leq \frac{\varepsilon-\varepsilon^\prime}{2}$. Since $\hat{y}\in \hV^{\varepsilon^\prime}_t(\o)$, there exists $\tilde{y}\in \tilde{\hV}^{\varepsilon^\prime}_t(\o)$ such that $|\hat{y}-\tilde{y}|\leq \varepsilon^\prime$. Hence,
\[
|y-\tilde{y}|\leq |y-\hat{y}|+|\hat{y}-\tilde{y}|\leq  \frac{\varepsilon-\varepsilon^\prime}{2}+\varepsilon^\prime\leq  \frac{\varepsilon+\varepsilon^\prime}{2}\leq \varepsilon.
\]
We also have $y\in \tilde{\hV}_t^{\varepsilon^\prime}(\o)\subseteq \tilde{\hV}_t^\varepsilon(\o)$. Thus, $\tilde{y}\in \hV_t^\varepsilon(\o)$.
\qed 

As an immediate consequence of Lemma~\ref{lem:V} and the definition in \eqref{eq:V}, we have
\bea\label{eq:clV}
\hV_t(\o)=\bigcap_{\varepsilon\in (0,1]}\cl(\hV_t^\varepsilon(\o)).
\eea
In particular, $\hV_t(\o)$ is a closed set.

Next, we formulate a simple inclusion that connects the functional $\varepsilon$-superhedging set $SHP^\varepsilon_t(X)$ and its random set analog $\cl(\hV_t^\varepsilon)$.

\begin{prop}\label{prop:SHPinV}
For every $\varepsilon\in [0,1]$, $t\in[0,T]$, it holds $\cl_{\hL^2}(SHP^\varepsilon_t(X))\subseteq \hS^2_{\cF_t}(\cl(\hV^\varepsilon_t))$.
\end{prop}

{\it Proof.} Since $\hS^2_{\cF_t}(\cl(\hV^\varepsilon_t))$ is a closed subset of $\hL^2_{\cF_t}(\hR^d)$, it is enough to show that $SHP^\varepsilon_t(X)\subseteq \hS^2_{\cF_t}(\cl(\hV^\varepsilon_t))$. Let $\eta\in SHP^{\varepsilon}_t(X)$. Hence, there exists $\hat{k}\in\hat{\hD}^2_{\hF}(\hR^d)$ such that $\hP(A|\cF_t)\geq 1-\varepsilon$, where
\[
A:=\cb{\hat{k}_r\in \hat{K}_r^{\varepsilon}\ \forall r\in[t,T],\ \eta-\int_t^T \hat{k}_rdr + L\varepsilon {\bf 1}\geq X}.
\]
Note that for $\hP$-a.e. $\o\in\O$,
\bea\label{eq:condP}
\hP(A|\cF_t)(\o)= \hP^{t,\o}(A^{t,\o})=\hP(A^{t,\o}).
\eea 
In particular, there exists $\O_0\in\cF$ with $\hP(\O_0)=1$ such that $\hP(A^{t,\o})\geq 1-\varepsilon$ for every $\o\in\O_0$.

Let us fix $\o\in\O_0$. Note that
\[
A^{t,\o}=\cb{\hat{k}^{t,\o}_r\in \hat{K}_r^{\varepsilon;t,\o}\ \forall r\in[t,T],\ \eta^{t,\o}-\int_t^T \hat{k}^{t,\o}_rdr + L\varepsilon {\bf 1}\geq X^{t,\o}}.
\]
Since $\eta$ is $\cF_t$-measurable, its shifted version $\eta^{t,\o}$ is deterministic, say, $\eta^{t,\o}=y$ $\hP$-a.s. for some $y\in\hR^d$.  Then, we have 
\[
\hP(A^{t,\o})=\hP\cb{\hat{k}^{t,\o}_r\in \hat{K}_r^{\varepsilon;t,\o}\ \forall r\in[t,T],\ y-\int_t^T \hat{k}^{t,\o}_rdr + L\varepsilon {\bf 1}\geq X^{t,\o}}\geq 1-\varepsilon.
\]
Note that $\hat{k}^{t,\o}\in \hat{\hD}^2_{\hF^t}(\hR^d)$.
Therefore, $y\in\tilde{\hV}^{\varepsilon}_t(\o)\subseteq \cl(\hV^{\varepsilon}_t(\o))$ so that $\eta^{t,\o}(\tilde{\o})\in\cl(\hV^{\varepsilon}_t(\o))$ for $\hP$-a.e. $\tilde{\o}\in \O$. Then, similar to \eqref{eq:condP} above, we have
\[
\hP\{\eta\in\cl(\hV^{\varepsilon}_t)\}=\int_{\O_0} \hP\{\eta^{t,\o}\in\cl(\hV_t^{\varepsilon;t,\o})\}\hP(d\o)=\int_{\O_0} \hP\{\eta^{t,\o}\in\cl(\hV^{\varepsilon}_t(\o))\}\hP(d\o)=1,
\]
where the second equality follows since $\cl(\hV^{\varepsilon}_t)$ is $\cF_t$-measurable, hence $\cl(\hV_t^{\varepsilon;t,\o})$ is deterministic for each $\o\in\O_0$. Hence, $\eta\in \hS^2_{\cF_t}(\cl(\hV^{\varepsilon}_t))$.
\qed 

In the next proposition, we show that the $\hL^2$-selectors of the local superhedging set $\hV_t$ can be calculated in terms of the $\hL^2$-selectors of its approximate versions.

\begin{prop}\label{prop:veps}
	For every $t\in[0,T]$, it holds
	\[
	\hS^2_{\cF_t}(\hV_t)=\bigcap_{\varepsilon\in(0,1]}\hS^2_{\cF_t}(\cl(\hV_t^{\varepsilon}))=\bigcap_{\varepsilon\in(0,1]}\of{\hS^2_{\cF_t}(\cl(\hV_t^{\varepsilon}))+\hB_{\hL^2}(\varepsilon)}.
	\]
	\end{prop}
	
{\it Proof.} The $\subseteq$ parts of both equalities are obvious. We first show that
\bea\label{eq:remaining1}
\bigcap_{\varepsilon\in(0,1]}\hS^2_{\cF_t}(\cl(\hV_t^{\varepsilon}))\subseteq \hS^2_{\cF_t}(\hV_t).
\eea
Let $\eta\in \bigcap_{\varepsilon\in(0,1]}\hS^2_{\cF_t}(\cl(\hV_t^{\varepsilon}))$. In particular, $\eta\in \hS^2_{\cF_t}(\cl(\hV_t^{\frac{1}{n}}))$ for every $n\in\hN$. By throwing away a $\hP$-null set for each $n\in\hN$, we see that $\eta \in \bigcap_{n\in\hN}(\cl(\hV_t^{\frac{1}{n}}))$ $\hP$-a.s. By Lemma~\ref{lem:V} and \eqref{eq:clV}, it is clear that this intersection equals $\hV_t$. Hence, $\eta\in\hS^2_{\cF_t}(\hV_t)$ and \eqref{eq:remaining1} follows.

Next, we show that
\bea\label{eq:remaining2}
\bigcap_{\varepsilon\in(0,1]}\of{\hS^2_{\cF_t}(\cl(\hV_t^{\varepsilon}))+\hB_{\hL^2}(\varepsilon)}\subseteq \bigcap_{\varepsilon\in(0,1]}\hS^2_{\cF_t}(\cl(\hV_t^{\varepsilon})).
\eea 
Let $\eta$ belong to the set on the left of \eqref{eq:remaining2}. Let us fix $\varepsilon\in(0,1]$. To show that $\eta\in \hS^2_{\cF_t}(\cl(\hV_t^{\varepsilon}))$, let us fix an arbitrary $\delta\in (0,2\varepsilon]$. Then, there exists $\tilde{\eta}\in\hS^2_{\cF_t}(\cl(\hV_t^{\frac{\delta}{2}}))$ such that
\[
\|\eta-\tilde{\eta}\|_2\leq \frac{\delta}{2}.
\]
Since $\tilde{\eta}\in\hS^2_{\cF_t}(\cl(\hV_t^{\frac{\delta}{2}}))$, by \cite[Theorem~1.3.31, Lemma~2.1.5]{Molchanov}, there exist $\eta^1,\ldots,\eta^n\in \hS^2_{\cF_t}(\hV_t^{\frac{\delta}{2}})$; $A_1,\ldots,A_n\in\cF_t$ partitioning $\O$; $n\in\hN$ such that
\[
\Big\|\tilde{\eta} - \sum_{i=1}^n \eta^i{\bf 1}_{A_i}\Big\|_2\leq \frac{\delta}{2}. 
\]
Then,
\[
\Big\|\eta - \sum_{i=1}^n \eta^i{\bf 1}_{A_i}\Big\|_2\leq \|\eta-\tilde{\eta}\|_2+\Big\|\tilde{\eta} - \sum_{i=1}^n \eta^i{\bf 1}_{A_i}\Big\|_2\leq \frac{\delta}{2}+\frac{\delta}{2}=\delta.
\]
Moreover, since $\delta\leq 2\varepsilon$, we have $\eta^i\in \hS^2_{\cF_t}(\hV_t^{\frac{\delta}{2}})\subseteq \hS^2_{\cF_t}(\hV_t^{\varepsilon})$ for each $i\in\{1,\ldots,n\}$. As $\delta\in (0,2\varepsilon]$ is arbitrary, by \cite[Theorem~1.3.31, Lemma~2.1.5]{Molchanov} again, we conclude that $\eta\in \hS^2_{\cF_t}(\cl(\hV_t^\varepsilon))$, proving \eqref{eq:remaining2}.
\qed 

\ms 
\no {\bf Functional vs. Pathwise Superhedging Sets.} The pathwise superhedging set $\hV_t$ can naively be considered as a local formulation of the functional superhedging set $SHP_t(X)$. The next theorem shows that this is not exactly correct; instead, the set of $\hL^2$-selectors of $\hV_t$ coincides with the ``approximation-closure" of $SHP_t(X)$, allowing an infinitesimal approximation error for the superhedging portfolios.

\begin{thm}\label{thm:V-SHP}
For every $t\in[0,T]$, it holds
\[
	\hS^2_{\cF_t}(\hV_t)= \bigcap_{\varepsilon\in (0,1]}\of{SHP_t^{\varepsilon}(X)+\hB_{\hL^2}(\varepsilon)}.
\]
\end{thm}

{\it Proof.} From Propositions~\ref{prop:SHPinV} and \ref{prop:veps}, we have
\[
\bigcap_{\varepsilon\in (0,1]}\of{SHP_t^{\varepsilon}(X)+\hB_{\hL^2}(\varepsilon)}\subseteq \bigcap_{\varepsilon\in (0,1]}\of{\hS^2_{\cF_t}(\cl(\hV_t^{\varepsilon}))+\hB_{\hL^2}(\varepsilon)}=\hS^2_{\cF_t}(\hV_t).
\]

To show the inclusion $\subseteq$ in the theorem, let $\eta \in \hS^2_{\cF_t}(\hV_t)$ and $\varepsilon\in (0,1]$. It is enough to find $\eta_{\varepsilon}\in SHP_t^{\varepsilon}(X)$ such that $\|\eta-\eta_{\varepsilon}\|_2\leq \varepsilon$.

By Proposition~\ref{prop:veps}, we have $\eta\in \hS^2_{\cF_t}(\cl(\hV_t^{\frac{\varepsilon}{2}}))$. Hence, by \cite[Theorem~1.3.31, Lemma~2.1.5]{Molchanov}, there exist $\eta^1,\ldots,\eta^n\in \hS^2_{\cF_t}(\hV_t^{\frac{\varepsilon}{2}})$; $A_1,\ldots,A_n\in\cF_t$ partitioning $\O$; $n\in\hN$ such that
\[
\Big\|\eta - \sum_{i=1}^n \eta^i{\bf 1}_{A_i}\Big\|_2\leq \frac{\varepsilon}{2}. 
\]
By throwing away finitely many null sets from $\O$, without loss of generality, we assume that $\eta(\o)\in\hV_t(\o)$ and $\eta^1(\o),\ldots,\eta^n(\o)\in \hV_t^{\frac{\varepsilon}{2}}(\o)$ for every $\o\in\O$.

We fix a Borel-measurable partition $(O_j)_{j\in\hN}$ of $\hR^d$ and a Borel-measurable partition $(C_\ell)_{\ell\in\hN}$ of $\hC_{s_0}([0,T],\hR^d)$ such that, for every $j,\ell\in\hN$,
\[
\sup_{y,y^\prime\in O_j}|y-y^\prime|\leq \frac{\varepsilon}{2},\quad \sup_{f^1,f^2\in C_\ell}\|f^1-f^2\|_\infty\leq \frac{\varepsilon\sqrt{\varepsilon}}{2\sqrt{2c}},
\]
where $c\geq 1$ is a constant to be defined later.

For $\theta:=(i,j,\ell)\in\Theta:=\{1,\ldots,n\}\times\hN\times\hN$, let us define
\bea\label{eq:partition}
B_\theta:=A_i\cap \{\eta^i\in O_j\}\cap \{S_{t\wedge \cdot}\in C_{\ell}\}.
\eea
Then, $(B_\theta)_{\theta\in \Theta}$ is an $\cF_t$-measurable partition of $\O$.

Let us fix $\theta=(i,j,\ell)\in\Theta$ and a representative outcome $\o^{\theta}\in B_\theta$. Since $\eta^i(\o^\theta)\in \hV_t^{\frac{\varepsilon}{2}}(\o^\theta)$, there exists $y^\theta\in \tilde{\hV}^{\frac{\varepsilon}{2}}_t(\o^\theta)$ such that
\bea\label{eq:yeta}
|y^\theta - \eta^i(\o^\theta)|\leq \frac{\varepsilon}{2}.
\eea 
Moreover, since $y^\theta\in \tilde{\hV}^{\frac{\varepsilon}{2}}_t(\o^\theta)$, there exists $\hat{k}^{\theta}\in\hat{\hD}^2_{\hF^t}(\hR^d)$ such that $\hP(F_{\theta})\geq 1-\frac{\varepsilon}{2}$, where
\bea\label{eq:ytheta}
F_\theta:=\cb{\hat{k}^{\theta}_r\in \hat{K}_r^{\frac{\varepsilon}{2};t,\o^\theta}\ \forall r\in[t,T],\ y^\theta-\int_t^T \hat{k}^{\theta}_rdr+L\frac{\varepsilon}{2}{\bf 1} \geq X^{t,\o^{\theta}}}.
\eea
Let
\[
S^{\Theta}_t:=\sum_{\theta\in\Theta}S_t(\o^{\theta}){\bf 1}_{B_\theta}.
\]
Then, by standard SDE estimates, there exists a constant $c\geq 1$ such that
\bea\label{eq:SDEest}
\hE\sqb{\sup_{r\in[t,T]}|S_r-S_r^{S_t^{\Theta};t}|^2\mid \cF_t}=\hE\sqb{\sup_{r\in[t,T]}|S_r^{S_t;t}-S_r^{S_t^{\Theta};t}|^2\mid \cF_t}\leq c|S_t-S_t^{\Theta}|^2\; \hP\text{-a.s.}\neg\neg\neg\neg\neg\neg\neg\neg\neg\neg\neg\neg\neg\neg\neg\neg\neg\neg
\eea
Let $\o\in B_\theta$. Similar to the derivation of \eqref{eq:shifted}, it is easy to observe that $(S^{S^{\Theta}_t;t})^{t,\o}$ has the same dynamics as in \eqref{eq:shifted} but with initial condition $(S_t^{S^{\Theta}_t;t})^{t,\o}=S_t(\o^\theta)$. Hence, by the uniqueness of the solution of the SDE, we have $(S^{S^{\Theta}_t;t})^{t,\o}=S^{t,\o^\theta}$ $\hP$-a.s. After using this with \eqref{eq:tomega} and \eqref{eq:partition}, \eqref{eq:SDEest} implies that
\[
\hE\sqb{\sup_{r\in[t,T]}|S_r^{t,\o}-S_r^{t,\o^{\theta}}|^2}\leq c|S_t(\o)-S_t(\o^\theta)|^2\leq c\frac{\varepsilon^3}{8c}=\frac{\varepsilon^3}{8}
\]
for $\hP$-a.e. $\o\in B_\theta$. Let us define 
\[
E:=\cb{\sup_{r\in[t,T]}|S_r-S_r^{S_t^{\Theta};t}|\leq \frac{\varepsilon}{2}}
\]
so that, whenever $\o\in B_{\theta}$, we have
\[
E^{t,\o}=\cb{\sup_{r\in[t,T]}|S_r^{t,\o}-S_r^{t,\o^{\theta}}|\leq \frac{\varepsilon}{2}}.
\]
Then, by Markov inequality, we have 
\[
1-\hP(E^{t,\o})=\hP\cb{\sup_{r\in[t,T]}|S_r^{t,\o}-S_r^{t,\o^{\theta}}|^2> \frac{\varepsilon^2}{4}}\leq \frac{\frac{\varepsilon^3}{8}}{\frac{\varepsilon^2}{4}}=\frac{\varepsilon}{2},
\]
and hence $\hP(E^{t,\o})\geq 1-\frac{\varepsilon}{2}$ for $\hP$-a.e. $\o\in B_\theta$.

Let $\o\in B_\theta$ be such that $\hP(E^{t,\o})\geq 1-\frac{\varepsilon}{2}$. Then, for every $\tilde{\o}\in \O$, we have
\beaa
|X^{t,\o}(\tilde{\o})-X^{t,\o^\theta}(\tilde{\o})|&=&|g(S^{t,\o}(\tilde{\o}))-g(S^{t,\o^\theta}(\tilde{\o}))|\\
&\leq& L\|S^{t,\o}(\tilde{\o})-S^{t,\o^\theta}(\tilde{\o})\|_\infty\\
&=& L\max\cb{\|S_{t\wedge \cdot}(\o)-S_{t\wedge\cdot}(\o^\theta)\|_\infty,\sup_{r\in[t,T]}|S^{t,\o}_r(\tilde{\o})-S^{t,\o^\theta}_r(\tilde{\o})|}\\
&\leq & L\max\cb{\frac{\varepsilon\sqrt{\varepsilon}}{2\sqrt{2c}},\sup_{r\in[t,T]}|S^{t,\o}_r(\tilde{\o})-S^{t,\o^\theta}_r(\tilde{\o})|}.
\eeaa
Let us fix $\tilde{\o}\in E^{t,\o}$. Then, we also have 
\bea\label{eq:xtheta}
|X^{t,\o}(\tilde{\o})-X^{t,\o^\theta}(\tilde{\o})|\leq L\max\cb{\frac{\varepsilon\sqrt{\varepsilon}}{2\sqrt{2c}},\frac{\varepsilon}{2}}=L\frac{\varepsilon}{2}\max\cb{\sqrt{\frac{\varepsilon}{2c}},1}=L\frac{\varepsilon}{2}
\eea
since we have $\varepsilon\leq 1\leq c\leq 2c$. 

Now, let us suppose further that $\tilde{\o}\in E^{t,\o}\cap F_\theta$. Then, using \eqref{eq:xtheta} in \eqref{eq:ytheta}, we obtain
\[
y^\theta-\int_t^T \hat{k}_r^{\theta}(\tilde{\o})dr+L\frac{\varepsilon}{2}{\bf 1}\geq X^{t,\o^\theta}(\tilde{\o})\geq X^{t,\o}(\tilde{\o})-L\frac{\varepsilon}{2}{\bf 1},
\]
which implies
\bea\label{eq:thetaconst}
y^\theta-\int_t^T \hat{k}_r^{\theta}(\tilde{\o})dr+L\varepsilon{\bf 1}\geq  X^{t,\o}(\tilde{\o}).
\eea
Moreover, by Lemma~\ref{lem:Keps}, having $\hat{k}^\theta_r(\tilde{\o})\in\hat{\sK}^{\frac{\varepsilon}{2}}(S_r^{t,\o^\theta}(\tilde{\o}))$ and $|S_r^{t,\o}(\tilde{\o})-S_r^{t,\o^\theta}(\tilde{\o})|\leq \frac{\varepsilon}{2}$, we conclude for each $r\in [t,T]$ that
\bea\label{eq:coneconst}
\hat{k}^\theta_r(\tilde{\o})\in \hat{\sK}^{\varepsilon}(S^{t,\o}_r(\tilde{\o}))=\hat{K}_r^{\varepsilon;t,\o}(\tilde{\o}).
\eea

Let us define
\[
\eta_{\varepsilon}(\o):=\sum_{\theta\in\Theta}y^{\theta}{\bf 1}_{B_\theta}(\o)
\]
and
\[
\hat{k}_r(\o):=\hat{k}^0_r(\o){\bf 1}_{[0,t)}(r)+{\bf 1}_{[t,T]}(r)\sum_{\theta\in\Theta}\hat{k}^{\theta}_r(\tilde{W}(\o)){\bf 1}_{B_\theta}(\o),
\]
where $\hat{k}^0\in \hat{\hD}_{\hF}^2(\hR^d)$ is arbitrarily fixed and $\tilde{W}\colon \O\to\O$ is defined by
\bea\label{eq:Wtilde}
\tilde{W}_u(\o):=(\o_{u+t}-\o_t){\bf 1}_{[0,T-t)}(u)+(\o_{T}-\o_t){\bf 1}_{[T-t,T]}(u),\quad u\in[0,T],
\eea
for each $\o\in\O$. Then, these definitions guarantee that $\hat{k}\in\hat{\hD}^2_{\hF}(\hR^d)$ and, for every $\tilde{\o}\in E^{t,\o}$ and $r\in[t,T]$, we have $\hat{k}^{t,\o}_r(\tilde{\o})=\hat{k}_r^\theta(\tilde{\o})\in \hat{K}_r^{\varepsilon;t,\o}(\tilde{\o})$ and $(\eta_\varepsilon)^{t,\o}(\tilde{\o})=y^\theta$. In particular, by \eqref{eq:thetaconst} and \eqref{eq:coneconst}, we get
\beaa
& &\hP\cb{\hat{k}^{t,\o}_r\in\hat{K}^{\varepsilon;t,\o}_r\ \forall r\in[t,T],\ (\eta_{\varepsilon})^{t,\o}-\int_t^T \hat{k}^{t,\o}_rdr+L\varepsilon{\bf 1}\geq X^{t,\o}}\\
& & \geq \hP(E^{t,\o}\cap F_\theta)\geq \hP(E^{t,\o})+\hP(F_\theta)-1 \geq \of{1-\frac{\varepsilon}{2}}+\of{1-\frac{\varepsilon}{2}}-1= 1-\varepsilon.
\eeaa
Equivalently, we have
\[
\hP\cb{\hat{k}_r\in\hat{K}^{\varepsilon}_r\ \forall r\in[t,T],\ \eta_\varepsilon-\int_t^T \hat{k}_r dr +L\varepsilon{\bf 1}\geq X\mid \cF_t}\geq 1-\varepsilon\quad \hP\text{-a.s.}
\]
Hence, $\eta_\varepsilon \in SHP^{\varepsilon}_t(X)$.

Finally, note that
\[
\norm{\eta-\eta_\varepsilon}_2\leq  \Big\|\eta-\sum_{i=1}^n \eta^i {\bf 1}_{A_i}\Big\|_2 +\Big\|\sum_{i=1}^n \eta^i {\bf 1}_{A_i}-\eta_\varepsilon\Big\|_2\leq  \frac{\varepsilon}{2}+\Big\|\sum_{\theta=(i,j,\ell)\in\Theta}(\eta^i-y^\theta){\bf 1}_{B_\theta}\Big\|_2 \leq \varepsilon 
\]
since, for $\o\in B_{\theta}$ with $\theta=(i,j,\ell)$, we have $|\eta^i(\o)-y^\theta|\leq |\eta^i(\o)-\eta^i(\o^\theta)|+|\eta^i(\o^\theta)-y^\theta|\leq \frac{\varepsilon}{2}$ by \eqref{eq:partition} and \eqref{eq:yeta}. Therefore, $\eta \in SHP^{\varepsilon}_t(X)+\hB_{\hL^2}(\varepsilon)$.
\qed

\ms 
\no {\bf The Pathwise Dynamic Programming Principle.} We begin with a lemma that will be used in the proof of the main result. It provides a simple one-sided concentration inequality for conditional expectations; we include its short proof for completeness.

\begin{lem}\label{lem:concentration}
	Let $\cG$ be a sub-$\sigma$-algebra of $\cF$. Let $A\in\cF$ and $\varepsilon 
	\in (0,1]$. Then, the following implication holds:
	\[
	\hP(A)\leq \varepsilon \quad \Rightarrow \quad \hP\{\hP(A|\cG)\geq \sqrt{\varepsilon}\}\leq \sqrt{\varepsilon}.
	\]
	\end{lem}
	
{\it Proof.} Let $Y\colon \O\to\hR_+$ be an $\cF$-measurable random variable such that $\hE[Y]\leq \varepsilon$. Let $\delta>0$. Then, we have $\varepsilon \geq \hE[Y]=\hE[Y{\bf 1}_{\{Y\geq \delta\}}]+\hE[Y{\bf 1}_{\{Y< \delta\}}]\geq \delta\hP\{Y\geq \delta\}$, which implies that $\hP\{Y\geq \delta\}\leq \frac{\varepsilon}{\delta}$. In particular, taking $\delta=\sqrt{\varepsilon}$ gives $\hP\{Y\geq \sqrt{\varepsilon}\} \leq \sqrt{\varepsilon}$. Finally, by taking $Y=\hP(A|\cG)$, the lemma follows since $\hE[Y]=\hP(A)$ by tower property.
\qed 

To formulate the pathwise dynamic programming principle, we generalize the concept of a superhedging portfolio for a subinterval of $[0,T]$ next.

\begin{defn}
Let $\o\in\O$, $0\leq t\leq u\leq T$, and $\varepsilon \in [0,1]$. We say that a vector $y\in\hR^d$ is an $\varepsilon$-superhedging portfolio at $\o$ over $[t,u]$ if there exist a strategy $\hat{k}^\o\in\hat{\hD}^2_{\hF^t}(\hR^d)$ and a target claim $\xi\in\hL^2_{\cF_u}(\hR^d)$ at time $u$ such that the following conditions are satisfied:

\ss
1. {\bf Approximate feasibility of the target:} It holds $\hP\{\xi\in \hV_u^{\varepsilon;t,\o}\}\geq 1-\varepsilon$.
	
\ss
2.  {\bf Approximate superhedging over $[t,u]$:} It holds
	\[
	\hP\cb{\hat{k}^\o_r\in \hat{K}_r^{\varepsilon;t,\o}\ \forall r\in[t,u],\ y-\int_t^u \hat{k}^\o_rdr +L\varepsilon{\bf 1}\geq \xi^{t,\o}}\geq 1-\varepsilon.
	\]

	We denote $\tilde{\hV}_{t,u}^{\varepsilon}(\o)$ to be the set of all $\varepsilon$-superhedging portfolios at $\o$ over $[t,u]$.
	\qed
	\end{defn}

The next theorem is the main result of this section. It relates the pathwise superhedging sets at different times through a \emph{set-valued dynamic programming principle}.

\begin{thm}\label{thm:DPP}
	Let $\o\in\O$ and $0\leq t\leq u\leq T$. Then, it holds
\[
\hV_t(\o)=\bigcap_{\varepsilon\in (0,1]}\of{\tilde{\hV}_{t,u}^\varepsilon(\o)+\hB_{\hR^d}(\varepsilon)}.
\]
	\end{thm}
	
{\it Proof:} To avoid dealing with complicated notation, we will prove the theorem for the case $t=0$ and $\o\equiv 0$:
\bea\label{eq:DPPsimple}
\hV_0(0)=\bigcap_{\varepsilon\in (0,1]}\of{\tilde{\hV}_{0,u}^\varepsilon(0)+\hB_{\hR^d}(\varepsilon)},
\eea
where $\tilde{\hV}^{\varepsilon}_{0,u}(0)$ simplifies as
\beaa
\tilde{\hV}^{\varepsilon}_{0,u}(0)=&\left\{y\in \hR^d\colon \ba{l}\exists \hat{k}\in \hat{\hD}^2_{\hF}(\hR^d),\ \exists\xi \in \hL^2_{\cF_u}(\hR^d),\quad \hP\{\xi\in \hV^{\varepsilon}_u\}\geq 1-\varepsilon, \\
\hP\cb{\hat{k}_r\in \hat{K}_r^{\varepsilon}\ \forall r\in[0,u],\ y-\int_0^u \hat{k}_rdr +L\varepsilon{\bf 1}\geq \xi}\geq 1-\varepsilon
\ea 
\right\}.
\eeaa 
The proof of the general case works similarly.

To prove the inclusion $\subseteq$ in \eqref{eq:DPPsimple}, let $y\in\hV_0(0)$ and fix $\varepsilon\in (0,1]$. We aim to find $\bar{y}\in \tilde{\hV}_{0,u}^{\varepsilon}(0)$ such that $|y-\bar{y}|\leq \varepsilon$.

Let $\delta\in (0,\varepsilon]$ to be chosen later depending on $\varepsilon$. By \eqref{eq:V}, there exists $y^{\delta}\in \tilde{\hV}_0^\delta(0)$ such that $|y-y^\delta|\leq \delta$. Since $y^\delta\in\tilde{\hV}_0^\delta(0)$, there exists $\hat{k}^\delta\in \hat{\hD}^2_{\hF}(\hR^d)$ such that
\bea\label{eq:kdelta}
\hP\cb{\hat{k}^\delta_r\in \hat{K}_r^\delta\ \forall r\in [0,T],\ y^\delta-\int_0^T \hat{k}_r^\delta dr+L\delta{\bf 1}\geq X}\geq 1-\delta.
\eea 
Let $\xi^\delta:=y^\delta-\int_0^u \hat{k}^\delta_rdr \in \hL^2_{\cF_u}(\hR^d)$. Obviously, we have
\[
y^\delta-\int_0^u \hat{k}_r^\delta dr + L\delta{\bf 1} \geq \xi^\delta\quad \hP\text{-a.s.}
\]
This, together with \eqref{eq:kdelta}, implies that
\[
\hP\cb{\hat{k}^\delta_r\in \hat{K}^\delta_r\ \forall r\in[0,u],\ y^\delta-\int_0^u \hat{k}_r^\delta dr +L\delta{\bf 1}\geq \xi^\delta}\geq 1-\delta.
\]
Since we assume that $\delta \leq \varepsilon$, we also have
\bea\label{eq:xidelta}
\hP\cb{\hat{k}^\delta_r\in \hat{K}^\varepsilon_r\ \forall r\in[0,u],\ y^\delta-\int_0^u \hat{k}_r^\delta dr +L\varepsilon{\bf 1}\geq \xi^\delta}\geq 1-\varepsilon.
\eea

Moreover, by \eqref{eq:kdelta}, we have $\hP(B)\geq 1-\delta$, where
\[
B:=\cb{\hat{k}^\delta_r\in \hat{K}^\delta_r\ \forall r\in [u,T],\ \xi^\delta-\int_u^T \hat{k}_r^\delta dr+L\delta {\bf 1}\geq X}.
\]
Then, by applying Lemma~\ref{lem:concentration} to the event $B^c$, we get $\hP\{\hP(B^c|\cF_u)\geq \sqrt{\delta}\}\leq \sqrt{\delta}$, or equivalently, $\hP\{\hP(B|\cF_u)>1-\sqrt{\delta}\}\geq 1-\sqrt{\delta}$. Hence, we choose $\delta=\varepsilon^2 \in (0,\varepsilon]$ and obtain 
\bea\label{eq:condprob-con}
\hP\{\hP(B|\cF_u)>1-\varepsilon\}\geq 1-\varepsilon.
\eea
Note that $\hP(B|\cF_u)(\o)=\hP(B^{u,\o})$ for $\hP$-a.e. $\o\in\O$, where
\[
B^{u,\o}=\cb{(\hat{k}_r^{\varepsilon^2})^{u,\o}\in\hat{K}_r^{\varepsilon^2;u,\o}\ \forall r\in[u,T],\ (\xi^{\varepsilon^2})^{u,\o}-\int_u^T (\hat{k}^{\varepsilon^2}_r)^{u,\o}dr +L\varepsilon^2{\bf 1}\geq X^{u,\o}}.
\]
Let $\bar{\O}:=\{\o\in\O\colon \xi^{\varepsilon^2}(\o)=\hE[(\xi^{\varepsilon^2})^{u,\o}]\}$. Since $\xi^{\varepsilon^2}$ is $\cF_u$-measurable, we have $\hP(\bar{\O})=1$. We claim that
\bea\label{eq:xi-claim}
\{\o\in\bar{\O}\colon \hP(B^{u,\o})\geq 1-\varepsilon\}\subseteq \{\o\in\bar{\O}\colon \xi^{\varepsilon^2}(\o)\in\hV_u^{\varepsilon}(\o)\}.
\eea
To see this, let $\o\in\bar{\O}$ be such that $\hP(B^{u,\o})\geq 1-\varepsilon$. Let us take $\hat{k}^\o:=(\hat{k}^{\varepsilon^2})^{u,\o}\in \hat{\hD}^2_{\hF^u}(\hR^d)$. Since $\xi^{\varepsilon^2}$ is $\cF_u$-measurable, $(\xi^{\varepsilon^2})^{u,\o}$ is deterministic $\hP$-a.s.; hence, $(\xi^{\varepsilon^2})^{u,\o}=\hE[(\xi^{\varepsilon^2})^{u,\o}]=\xi^{\varepsilon^2}(\o)$ $\hP$-a.s. because $\o\in\bar{\O}$. Let $y^\o:=\xi^{\varepsilon^2}(\o)$. Then, since $\hP(B^{u,\o})\geq 1-\varepsilon$, we have
\beaa
&&\hP\cb{\hat{k}^\o_r\in \hat{K}_r^{\varepsilon;u,\o}\ \forall r\in[u,T],\ y^\o-\int_u^T \hat{k}^\o_rdr+L\varepsilon{\bf 1}\geq X^{u,\o}}\\
&&\geq \hP\cb{\hat{k}^\o_r\in \hat{K}_r^{\varepsilon^2;u,\o}\ \forall r\in[u,T],\ y^\o-\int_u^T \hat{k}^\o_rdr+L\varepsilon^2{\bf 1}\geq X^{u,\o}}\geq 1-\varepsilon.
\eeaa
Hence, $\xi^{\varepsilon^2}(\o)=y^\o\in \tilde{\hV}_u^{\varepsilon}(\o)\subseteq \hV_u^{\varepsilon}(\o)$, which finishes the proof of \eqref{eq:xi-claim}.

From \eqref{eq:xi-claim} and \eqref{eq:condprob-con}, we get $\hP\{\xi^{\varepsilon^2}\in \hV_u^\varepsilon\}\geq 1-\varepsilon$. By combining this with \eqref{eq:xidelta} and taking $\bar{y}=y^{\varepsilon^2}$, we see that $\bar{y}\in\tilde{\hV}_{0,u}^{\varepsilon}(0)$ and $|y-\bar{y}|\leq\varepsilon$; hence, $y\in \tilde{\hV}^\varepsilon_{0,u}(0)+\hB_{\hR^d}(\varepsilon)$, completing the proof of the inclusion $\subseteq$ in \eqref{eq:DPPsimple}.

To prove the inclusion $\supseteq$ in \eqref{eq:DPPsimple}, we follow a similar line of thought as in the proof of Theorem~\ref{thm:V-SHP} using partitioning ideas and SDE estimates. To that end, let $y\in\hR^d$ be such that $y\in \tilde{\hV}^\delta_{0,u}(0)+\hB_{\hR^d}(\delta)$ for every $\delta\in (0,1]$. Let us fix $\varepsilon\in (0,1]$. Thanks to \eqref{eq:V}, we aim to show that $y\in \tilde{\hV}^\varepsilon_0(0)+\hB_{\hR^d}(\varepsilon)$.

By the choice of $y$, taking $\delta=\frac{\varepsilon}{4}$ yields the existence of $\tilde{y}\in \tilde{\hV}_{0,u}^{\frac{\varepsilon}{4}}(0)$ such that $|y-\tilde{y}|\leq \frac{\varepsilon}{4}$. Since $\tilde{y}\in\tilde{\hV}^{\frac{\varepsilon}{4}}_{0,u}(0)$, there exist $\hat{k}\in \hat{\hD}^2_{\hF}(\hR^d)$, $\xi\in\hL^2_{\cF_u}(\hR^d)$ such that $\hP\{\xi\in \hV_u^{\frac{\varepsilon}{4}}\}\geq 1-\frac{\varepsilon}{4}$ and
\bea\label{eq:C1prob}
\hP\cb{\hat{k}_r\in \hat{K}_r^{\frac{\varepsilon}{4}}\ \forall r\in[0,u],\ \tilde{y}-\int_0^u \hat{k}_r dr +L\frac{\varepsilon}{4}{\bf 1}\geq \xi}\geq 1-\frac{\varepsilon}{4}.
\eea
As justified above, without loss of generality, we assume that $\xi^{u,\o}=\xi(\o)$ a.s. for every $\o\in\O$.

Let us fix a Borel-measurable partition $(O_j)_{j\in\hN}$ of $\hR^d$ and a Borel-measurable partition $(C_\ell)_{\ell\in\hN}$ of $\hC_{s_0}([0,T],\hR^d)$ such that, for every $j,\ell\in\hN$,
\[
\sup_{y^{\prime},y^{\prime\prime}\in O_j}|y^{\prime}-y^{\prime\prime}|\leq \frac{\varepsilon}{8},\quad \sup_{f^1,f^2\in C_\ell}\|f^1-f^2\|_\infty\leq \frac{\varepsilon\sqrt{\varepsilon}}{16\sqrt{c}},
\]
where $c\geq 1$ is a constant to be defined later.

Let $B:=\{\xi\in \hV_u^{\frac{\varepsilon}{4}}\}\in\cF_u$. For each $(j,\ell)\in\hN^2$, let us define
\bea\label{eq:partitionDPP}
B_{j,\ell}:=B\cap \{\xi \in O_j\}\cap \{S_{u\wedge \cdot}\in C_{\ell}\}.
\eea
Then, $(B_{j,\ell})_{(j,\ell)\in \hN^2}$ is an $\cF_u$-measurable partition of $B$.

Let us fix $(j,\ell)\in\hN^2$ and a representative outcome $\o^{j,\ell}\in B_{j,\ell}$. Since $\xi(\o^{j,\ell})\in \hV_u^{\frac{\varepsilon}{4}}(\o^{j,\ell})$, there exists $y^{j,\ell}\in \tilde{\hV}_u^{\frac{\varepsilon}{4}}(\o^{j,\ell})$ such that
\bea\label{eq:y-xi}
|y^{j,\ell}-\xi(\o^{j,\ell})|\leq \frac{\varepsilon}{4}.
\eea
Moreover, since $y^{j,\ell}\in \tilde{\hV}_u^{\frac{\varepsilon}{4}}(\o^{j,\ell})$, there exists $\hat{k}^{j,\ell}\in \hat{\hD}^2_{\hF^u}(\hR^d)$ such that $\hP(F_{j,\ell})\geq 1-\frac{\varepsilon}{4}$, where
\bea\label{eq:ythetaDPP}
F_{j,\ell}:=\cb{\hat{k}^{j,\ell}_r\in \hat{K}^{\frac{\varepsilon}{4};u,\o^{j,\ell}}_r\ \forall r\in [u,T],\ y^{j,\ell}-\int_u^T \hat{k}^{j,\ell}_rdr +L\frac{\varepsilon}{4}{\bf 1}\geq X^{u,\o^{j,\ell}}}.
\eea

Let us also fix a representative outcome $\o^0\in B^c$. Let
\[
\bar{S}_u:=\sum_{(j,\ell)\in\hN^2}S_u(\o^{j,\ell}){\bf 1}_{B_{j,\ell}}+S_u(\o^0){\bf 1}_{B^c}.
\]
Then, by standard SDE estimates, there exists a constant $c\geq 1$ such that
\bea\label{eq:SDEestDPP}
\hE\sqb{\sup_{r\in[u,T]}\neg |S_r-S_r^{\bar{S}_u;u}|^2\neg \mid\neg \cF_u\neg }\neg=\neg\hE\sqb{\sup_{r\in[u,T]}\neg |S_r^{S_u;u}-S_r^{\bar{S}_u;u}|^2\neg \mid \neg\cF_u\neg }\neg \leq \neg c|S_u-\bar{S}_u|^2\; \hP\text{-a.s.} \neg\neg\neg\neg\neg\neg\neg\neg\neg\neg\neg\neg\neg
\eea
Let $\o\in B_{j,\ell}$. Repeating the arguments in the proof of Theorem~\ref{thm:V-SHP} together with \eqref{eq:partitionDPP}, \eqref{eq:SDEestDPP}, and Markov inequality, we obtain $\hP(E^{u,\o})\geq 1-\frac{\varepsilon}{4}$ for $\hP$-a.e. $\o\in B_{j,\ell}$, where
\[
E^{u,\o}:=\cb{\sup_{r\in[u,T]}|S_r^{u,\o}-S_r^{u,\o^{j,\ell}}|\leq \frac{\varepsilon}{8}}.
\]

Let $\o\in B_{j,\ell}$ be such that $\hP(E^{u,\o})\geq 1-\frac{\varepsilon}{4}$. Similar to \eqref{eq:xtheta} in the proof of Theorem~\ref{thm:V-SHP}, for every $\tilde{\o}\in E^{u,\o}$, we have 
\bea\label{eq:xthetaDPP}
|X^{t,\o}(\tilde{\o})-X^{t,\o^{j,\ell}}(\tilde{\o})|\leq L\frac{\varepsilon}{8}.
\eea
Now, let us suppose further that $\tilde{\o}\in E^{u,\o}\cap F_{j,\ell}$ and $\xi^{u,\o}(\tilde{\o})=\xi(\o)$. Then, using \eqref{eq:xthetaDPP} in \eqref{eq:ythetaDPP}, we obtain
\[
y^{j,\ell}-\int_u^T \hat{k}_r^{j,\ell}(\tilde{\o})dr+L\frac{\varepsilon}{4}{\bf 1}\geq X^{u,\o^{j,\ell}}(\tilde{\o})\geq X^{u,\o}(\tilde{\o})-L\frac{\varepsilon}{8}{\bf 1},
\]
which implies
\bea\label{eq:thetaconstDPP}
y^{j,\ell}-\int_u^T \hat{k}_r^{j,\ell}(\tilde{\o})dr+L\frac{3\varepsilon}{8}{\bf 1}\geq  X^{u,\o}(\tilde{\o}).
\eea
By \eqref{eq:partitionDPP} and \eqref{eq:y-xi}, we also have
\[
|\xi^{u,\o}(\tilde{\o})-y^{j,\ell}|=|\xi(\o)-y^{j,\ell}|\leq |\xi(\o)-\xi(\o^{j,\ell})|+|\xi(\o^{j,\ell})-y^{j,\ell}|\leq \frac{\varepsilon}{8}+\frac{\varepsilon}{4}=\frac{3\varepsilon}{8}.
\]
Combining this with \eqref{eq:thetaconstDPP} and recalling that $L\geq 1$ yield
\bea\label{eq:thetaconstDPP2}
\xi^{u,\o}(\tilde{\o})-\int_u^T \hat{k}_r^{j,\ell}(\tilde{\o})dr+L\frac{3\varepsilon}{4}{\bf 1}\geq y^{j,\ell}-\int_u^T \hat{k}_r^{j,\ell}(\tilde{\o})dr+L\frac{3\varepsilon}{8}{\bf 1}\geq  X^{u,\o}(\tilde{\o}).\neg\neg\neg
\eea

Moreover, for each $r\in [u,T]$, by Lemma~\ref{lem:Keps}, having $\hat{k}_r^{j,\ell}(\tilde{\o})\in\hat{\sK}^{\frac{\varepsilon}{4}}(S_r^{u,\o^{j,\ell}}(\tilde{\o}))$ and $|S_r^{u,\o}(\tilde{\o})-S_r^{u,\o^{j,\ell}}(\tilde{\o})|\leq \frac{\varepsilon}{8}$ implies that
\bea\label{eq:coneconstDPP}
\hat{k}_r^{j,\ell}(\tilde{\o})\in \hat{\sK}^{\frac{3\varepsilon}{8}}(S^{u,\o}_r(\tilde{\o}))=\hat{K}_r^{\frac{3\varepsilon}{8};u,\o}(\tilde{\o})\subseteq \hat{K}_r^{\varepsilon;u,\o}(\tilde{\o}).
\eea

For each $(r,\o)\in[0,T]\times\O$, let us define
\[
\hat{k}^{\varepsilon}_r(\o)\neg:=\neg\hat{k}_r(\o){\bf 1}_{B}(\o){\bf 1}_{[0,u)}(r)\neg+\neg \sum_{(j,\ell)\in\hN^2}\hat{k}^{j,\ell}_r(\tilde{W}(\o)){\bf 1}_{B_{j,\ell}}(\o){\bf 1}_{[u,T]}(r)\neg+\neg \hat{k}^0_r(\o){\bf 1}_{B^c}(\o){\bf 1}_{[0,T]}(r),
\]
where $\hat{k}^0\in \hat{\hD}_{\hF}^2(\hR^d)$ is arbitrarily fixed and $\tilde{W}\colon \O\to\O$ is defined by \eqref{eq:Wtilde}. This definition guarantees that $\hat{k}^{\varepsilon}\in\hat{\hD}^2_{\hF}(\hR^d)$.

To check that $\tilde{y} \in \tilde{\hV}^{\varepsilon}_0(0)$, it is enough to show that $\hP(C)\geq 1-\varepsilon$, where
\[
C:=\cb{\hat{k}_r^\varepsilon\in \hat{K}^\varepsilon_r\ \forall r\in[0,T],\ \tilde{y}-\int_0^T\hat{k}^{\varepsilon}_r dr +L\varepsilon {\bf 1}\geq X}.
\]
Clearly, we have $C\supseteq B\cap C_1\cap C_2$, where
\beaa
C_1& :=& \cb{\hat{k}_r^\varepsilon\in \hat{K}^\varepsilon_r\ \forall r\in[0,u],\ \tilde{y}-\int_0^u\hat{k}^{\varepsilon}_r dr +L\frac{\varepsilon}{4} {\bf 1}\geq \xi},\\
C_2&:=& \cb{\hat{k}_r^\varepsilon\in \hat{K}^\varepsilon_r\ \forall r\in[u,T],\ \xi-\int_u^T\hat{k}^{\varepsilon}_r dr +L\frac{3\varepsilon}{4} {\bf 1}\geq X}.
\eeaa
Recall that $B=\{\xi\in \hV_u^{\frac{\varepsilon}{4}}\}$. By the definition of $\hat{k}^\varepsilon$ and \eqref{eq:C1prob}, we have
\beaa
\hP(B\cap C_1)&\geq& \hP\of{B\cap \cb{\hat{k}_r\in \hat{K}^{\frac{\varepsilon}{4}}_r\ \forall r\in[0,u],\ \tilde{y}-\int_0^u\hat{k}_r dr +L\frac{\varepsilon}{4} {\bf 1}\geq \xi}} \\
&\geq & \hP(B)+\hP\cb{\hat{k}_r\in \hat{K}^{\frac{\varepsilon}{4}}_r\ \forall r\in[0,u],\ \tilde{y}-\int_0^u\hat{k}_r dr +L\frac{\varepsilon}{4} {\bf 1}\geq \xi}-1\geq 1-\frac{\varepsilon}{2}.
\eeaa 
Let us fix $(j,\ell)\in\hN^2$ and $\o\in B_{j,\ell}$ such that $\hP(E^{u,\o})\geq 1-\frac{\varepsilon}{4}$. The definition of $\hat{k}^\varepsilon$ yields that $(\hat{k}^\varepsilon)^{u,\o}=\hat{k}^{j,\ell}$. Hence, by \eqref{eq:thetaconstDPP2} and \eqref{eq:coneconstDPP}, we have
\beaa
\hP(C_2^{u,\o})&=&\hP\cb{\hat{k}_r^{j,\ell}\in \hat{K}^{\varepsilon;u,\o}_r\ \forall r\in[u,T],\ \xi^{u,\o}-\int_u^T\hat{k}^{j,\ell}_r dr +L\frac{3\varepsilon}{4} {\bf 1}\geq X^{u,\o}}\\
&\geq& \hP(E^{u,\o}\cap F_{j,\ell})\geq \hP(E^{u,\o})+\hP(F_{j,\ell})-1\geq 1-\frac{\varepsilon}{2}.
\eeaa 
Since $B\cap C_1\in \cF_u$, we have
\beaa 
\hP(C)&\geq &\hP(B\cap C_1\cap C_2)=\hE[{\bf 1}_{B\cap C_1}\hP(C_2|\cF_u)]=\int_{B\cap C_1}\hP(C_2^{u,\o})\hP(d\o)\\
&\geq & \of{1-\frac{\varepsilon}{2}}\hP(B\cap C_1)\geq \of{1-\frac{\varepsilon}{2}}^2=1-\varepsilon+\frac{\varepsilon^2}{4}\geq 1-\varepsilon.
\eeaa 
Therefore, $\tilde{y} \in \tilde{\hV}^{\varepsilon}_0(0)$. Since we also have $|y-\tilde{y}|\leq \frac{\varepsilon}{4}\leq \varepsilon$, this finishes the proof of the inclusion $\supseteq$ in \eqref{eq:DPPsimple}.
\qed 

 \bibliography{ACM2025arXiv}

@article{kovacova,
	title={Time consistency of the mean-risk problem},
	author={Kov{\'a}{\v{c}}ov{\'a}, Gabriela and Rudloff, Birgit},
	journal={Operations Research},
	volume={69},
	number={4},
	pages={1100--1117},
	year={2021}
}

@article{karnam2017dynamic,
	title={Dynamic approaches for some time-inconsistent optimization problems},
	author={Karnam, Chandrasekhar and Ma, Jin and Zhang, Jianfeng},
	journal={The Annals of Applied Probability},
	volume={27},
	number={6},
	pages={3435--3477},
	year={2017}
}

@article{dynamicgames,
	title = {Dynamic set values for nonzero sum games with multiple equilibriums},
	author = {Feinstein, Zachary and Rudloff, Birgit and Zhang, Jianfeng},
	year = {2022},
	journal = {Mathematics of Operations Research},
	volume = {47},
	number = {1},
	pages = {616--642}
}

@unpublished{hjb,
	title = {Set-valued {H}amilton-{J}acobi-{B}ellman equations},
	author = {İ\c{s}eri, Melih and Zhang, Jianfeng},
	note = {arXiv e-print 2311.05727},
	year = {2025}
}

@article{meanfield,
	title = {Set values for mean field games},
	author = {İ\c{s}eri, Melih and Zhang, Jianfeng},
	journal = {Transactions of the American Mathematical Society},
	volume = {377},
	number = {10},
	year = {2024},
	pages = {7117--7174}
}

@article{amw,
	author = {Ararat, {\c{C}}a\u{g}{\i}n and Ma, Jin and Wu, Wenqian},
	title = {Set-valued backward stochastic differential equations},
	journal = {The Annals of Applied Probability},
	volume = {33},
	number = {5},
	pages = {3418--3448},
	year = {2023}
}

@article{araratfeinstein,
	author = {Ararat, {\c{C}}a\u{g}{\i}n and Feinstein, Zachary},
	title = {Set-valued risk measures as backward stochastic difference inclusions and equations},
	journal = {Finance and Stochastics},
	volume = {25},
	number = {1},
	pages = {43--76},
	year = {2021}
}

@article{lohnerudloff,
	author = {L\"{o}hne, Andreas and Rudloff, Birgit},
	title = {An algorithm for calculating the set of superhedging portfolios in markets with transaction costs},
	journal = {International Journal of Theoretical and Applied Finance},
	volume = {17},
	number = {2},
	pages = {1450012},
	year = {2014}
}

@book{karatzas,
	author = {Karatzas, Ioannis and Shreve, Steven E.},
	title = {Brownian Motion and Stochastic Calculus},
	year = {1998},
	edition = {second},
	publisher = {Springer}
}

@article{LepMol,
	author = {L{\'e}pinette, Emmauel and Molchanov, Ilya},
	journal = {Journal of Mathematical Analysis and Applications},
	pages = {368-392},
	title = {Conditional cores and conditional convex hulls of random sets},
	volume = {478},
	number = {2},
	pages = {368--392},
	year = {2019}
}

@article{AlMa,
	author = {Almuzaini, Atiqah and Ma, Jin},
	journal = {Set-Valued and Variational Analysis},
	title = {Set-valued stochastic differential equations with unbounded coefficients},
	volume = {33},
	number = {2},
	pages = {15},
	year = {2025}
}

@article{cvi,
	author = {Cvitani{\'c}, Jak{\v{s}}a and Karatzas, Ioannis},
	journal = {Mathematical Finance},
	number = {2},
	pages = {133--165},
	title = {Hedging and portfolio optimization under transaction costs: a martingale approach},
	volume = {6},
	year = {1996}
}

@article{FR13,
	author = {Feinstein, Zachary and Rudloff, Birgit},
	journal = {Quantitative Finance},
	number = {9},
	pages = {1473--1489},
	title = {Time consistency of dynamic risk measures in markets with transaction costs},
	volume = {13},
	year = {2013}
}

@article{FR2015,
	author = {Feinstein, Zachary and Rudloff, Birgit},
	journal = {Finance and Stochastics},
	number = {1},
	pages = {67--107},
	publisher = {Springer},
	title = {Multi-portfolio time consistency for set-valued convex and coherent risk measures},
	volume = {19},
	year = {2015}
}

@article{jouini,
	author = {Jouini, Elyes and Meddeb, Moncef and Touzi, Nizar},
	journal = {Finance and Stochastics},
	number = {4},
	pages = {531--552},
	title = {Vector-valued coherent risk measures},
	volume = {8},
	year = {2004}
}

@article{Campi,
	author = {Campi, Luciano and Schachermayer, Walter},
	title = {A super-replication theorem in {K}abanov's model of transaction costs},
	journal = {Finance and Stochastics},
	volume = {10},
	number = {4},
	pages = {579--596},
	year = {2006}
}

@book{KabanovSafarian,
	author = {Kabanov, Yuri and Safarian, Mher},
	title = {Markets with Transaction Costs: Mathematical Theory},
	publisher = {Springer},
	year = {2009}
}

@article{KabanovRasonyi,
   author = {Kabanov, Yuri M and R\'{a}sonyi, Mikl\'{o}s and Stricker, Christophe},
   title = {No-arbitrage criteria for financial markets with efficient friction},
   journal = {Finance and Stochastics},
   volume = {6},
   number = {3},
   pages = {371--382},
   year = {2002}
}

@article{Kcurrency,
	author = {Kabanov, Yuri M},
	journal = {Finance and Stochastics},
	pages = {237--248},
	title = {Hedging and liquidation under transaction costs in currency markets},
	volume = {3},
	number = {2},
	year = {1999}
}

@article{KandG,
	author = {Kabanov, Yuri M and Last, G{\"u}nter},
	journal = {Mathematical Finance},
	number = {1},
	pages = {63--70},
	publisher = {Wiley Online Library},
	title = {Hedging under transaction costs in currency markets: A continuous-time model},
	volume = {12},
	year = {2002}
}

@incollection{KandS,
	author = {Kabanov, Yuri M and Stricker, Christophe},
	booktitle = {Advances in Finance and Stochastics: Essays in Honour of Dieter Sondermann},
	editor = {Sandmann, Klaus and Sch\"{o}nbucher, Philipp J.},
	pages = {125--136},
	publisher = {Springer},
	title = {Hedging of contingent claims under transaction costs},
	year = {2002}
}

@book{K,
	author = {Kisielewicz, Micha{\l}},
	publisher = {Springer},
	title = {Stochastic Differential Inclusions and Applications},
	year = {2013}
}

@book{MK,
	author = {Kisielewicz, Micha{\l}},
	publisher = {Springer},
	title = {Set-Valued Stochastic Integrals and Applications},
	year = {2020}
}

@book{Molchanov,
	author = {Molchanov, Ilya},
	edition = {second},
	publisher = {Springer},
	title = {Theory of Random Sets},
	year = {2017}
}

@book{rockafellar,
	author = {Rockafellar, R Tyrell},
	title = {Convex Analysis},
	year = {1970},
	publisher = {Princeton University Press}
}

@article{Sch,
	author = {Schachermayer, Walter},
	journal = {Mathematical Finance},
	number = {1},
	pages = {19--48},
	title = {The fundamental theorem of asset pricing under proportional transaction costs in finite discrete time},
	volume = {14},
	year = {2004}
}
\bibliographystyle{abbrv}

\end{document}